\documentclass[usenatbib,useAMS]{mn2e}
\usepackage{amssymb, amsmath, epsf, graphicx}
\usepackage{txfonts}
\usepackage{mathrsfs}
\usepackage{amsfonts}

\def\jnl@style{\it}
\def\aaref@jnl#1{{\jnl@style#1}}

\def\aaref@jnl#1{{\jnl@style#1}}

\def\aj{\aaref@jnl{AJ}}                   
\def\araa{\aaref@jnl{ARA\&A}}             
\def\apj{\aaref@jnl{ApJ}}                 
\def\apjl{\aaref@jnl{ApJ}}                
\def\apjs{\aaref@jnl{ApJS}}               
\def\ao{\aaref@jnl{Appl.~Opt.}}           
\def\apss{\aaref@jnl{Ap\&SS}}             
\def\aap{\aaref@jnl{A\&A}}                
\def\aapr{\aaref@jnl{A\&A~Rev.}}          
\def\aaps{\aaref@jnl{A\&AS}}              
\def\azh{\aaref@jnl{AZh}}                 
\def\baas{\aaref@jnl{BAAS}}               
\def\jrasc{\aaref@jnl{JRASC}}             
\def\memras{\aaref@jnl{MmRAS}}            
\def\mnras{\aaref@jnl{MNRAS}}             
\def\pra{\aaref@jnl{Phys.~Rev.~A}}        
\def\prb{\aaref@jnl{Phys.~Rev.~B}}        
\def\prc{\aaref@jnl{Phys.~Rev.~C}}        
\def\prd{\aaref@jnl{Phys.~Rev.~D}}        
\def\pre{\aaref@jnl{Phys.~Rev.~E}}        
\def\prl{\aaref@jnl{Phys.~Rev.~Lett.}}    
\def\pasp{\aaref@jnl{PASP}}               
\def\pasj{\aaref@jnl{PASJ}}               
\def\qjras{\aaref@jnl{QJRAS}}             
\def\skytel{\aaref@jnl{S\&T}}             
\def\solphys{\aaref@jnl{Sol.~Phys.}}      
\def\sovast{\aaref@jnl{Soviet~Ast.}}      
\def\ssr{\aaref@jnl{Space~Sci.~Rev.}}     
\def\zap{\aaref@jnl{ZAp}}                 
\def\nat{\aaref@jnl{Nature}}              
\def\iaucirc{\aaref@jnl{IAU~Circ.}}       
\def\aplett{\aaref@jnl{Astrophys.~Lett.}} 
\def\apspr{\aaref@jnl{Astrophys.~Space~Phys.~Res.}}
\def\bain{\aaref@jnl{Bull.~Astron.~Inst.~Netherlands}} 
\def\fcp{\aaref@jnl{Fund.~Cosmic~Phys.}}  
\def\gca{\aaref@jnl{Geochim.~Cosmochim.~Acta}}   
\def\grl{\aaref@jnl{Geophys.~Res.~Lett.}} 
\def\jcp{\aaref@jnl{J.~Chem.~Phys.}}      
\def\jgr{\aaref@jnl{J.~Geophys.~Res.}}    
\def\jqsrt{\aaref@jnl{J.~Quant.~Spec.~Radiat.~Transf.}}
\def\memsai{\aaref@jnl{Mem.~Soc.~Astron.~Italiana}}
\def\nphysa{\aaref@jnl{Nucl.~Phys.~A}}   
\def\physrep{\aaref@jnl{Phys.~Rep.}}   
\def\physscr{\aaref@jnl{Phys.~Scr}}   
\def\planss{\aaref@jnl{Planet.~Space~Sci.}}   
\def\procspie{\aaref@jnl{Proc.~SPIE}}   

\newcommand{\ergs}{erg~s$^{-1}$}
\newcommand\ion[2]{\mbox{#1{\thinspace\scshape #2}}}%

\begin{document}
\title[Type 1 Active Galactic Nuclei Fraction]
{Type 1 Active Galactic Nucleus Fraction in SDSS/FIRST Survey}

\author[Lu et al.]
  {Yu Lu, Ting-Gui Wang, Xiao-Bo Dong, Hong-Yan Zhou\\
  $^1$CAS Key Laboratory for Research in Galaxies and Cosmology, University of Science and Technology of
  China,\\
  $^2$ Center for Astrophysics, University of Science and Technology of China, Hefei, Anhui, 230026, P.R.China \\
Hefei, Anhui, 230026, China\\}

\maketitle

\begin{abstract}
\noindent
 In the unification scheme, narrow lined (type 2) active galactic nuclei
(AGN) are intrinsically similar to broad lined (type 1) AGN with the
exception that the line of sight to the broad emission line region
and accretion disk is blocked by a dusty torus. The fraction of type
1 AGN measures the average covering factor of the torus. In this
paper, we explore the dependence of this fraction on nuclear
properties for a sample of low redshift ($z \le 0.35$) radio strong
($P_{\rm{1.4GHz}}\ge 10^{23}$W Hz$^{-1}$) AGN selected by matching
the spectroscopic catalog of Sloan Digital Sky Survey and the radio
source catalog of Faint Image of Radio Sky at Twenty cm. After
correcting for several selection effects, we find that : (1) type 1
fraction $f_1$ keeps at a constant of $\sim20\%$ in the
[\ion{O}{iii}] luminosity range of
$40.7<\log(L_{\rm{[OIII]}}\rm{/erg\ s^{-1}})<43.5$. This result is
significantly different from previous studies, and the difference
can be explained by extinction correction and different treatment of
selection effects. (2) $f_1$ rises with black hole mass from
$\sim20\%$ below $10^8\rm{M}_{\sun}$ to ~30\% above that. This
coincides with the decrease of the fraction of highly-inclined disk
galaxies with black hole mass, implying a population of Seyfert
galaxies seen as Type-2 due to galaxy-scale obscuration in disk when
the host galaxy type transfer from bulge-dominant to disk-dominant.
 (3) $f_1$ is independent of the Eddington ratio for its value
between 0.01 and 1; (4) $f_1$ ascends from 15\% to 30\% in the radio
power range of $23<\log(P_{\rm{1.4GHz}}/\rm{W Hz^{-1}})<24$, then
remain a constant at $\sim 30\%$ up to $10^{26}$ W Hz$^{-1}$.

\end{abstract}


\footnotetext{$^{\star}$Email: laolu@mail.ustc.edu.cn}

\section{Introduction}
   Active Galactic Nuclei (AGN) are traditionally divided
into 2 subclasses: type 2 and type 1, according to the absence or
presence of the broad emission lines. Their different observed
properties can be explained largely via anisotropic obscuration,
likely by a dusty torus on scales of several to tens parsecs, of
otherwise the same type of objects (Antonucci 1993; Tran 1995;
 Nenkova et al. 2002). Because of its large
physical scale, Narrow emission Line Region (NLR) cannot be
(completely) obscured by such torus. Therefore, an AGN will appear
as a type 2 source, when our line of sight to the Broad emission
Line Region (BLR) and the accretion disk is blocked by the dusty
torus. Scattered broad lines have been detected in the polarized
light in almost half of type-2 AGN, which strongly supports the
unification scheme (Antonucci \& Miller 1985; Miller \& Goodrich
1990; Moran et al. 2000).

   An important parameter in this model is the opening angle of the
torus. The fraction of type 1 AGN (hereafter $f_1$) is a measure of
the average opening angle of torus in such a unification scheme.
Previous studies have shown a type 2 to type 1 ratio of 3:1 for
local Seyfert galaxies (e.g., Maia et al. 2003; c.f, Ho et al.
1997). However, there is no reason that the torus opening angle
should be the same for every AGN, rather it may depend on the black
hole mass, accretion rate, luminosity or other intrinsic parameters.
Exploring these parameters dependence will be an important extension
to the simple unification scheme, and yield the insight into the
origin of the dust torus as well.

The luminosities of narrow emission lines, such as [\ion{O}{iii}],
mid-infrared light and hard X-rays have been considered as isotropic
properties (e.g. Meisenheimer et al.2001; Kauffmann et al. 2003;
Wang et al. 2006). Thus, we can examine the luminosity dependence of
$f_1$ using those luminosities. It was found that $f_1$ increases
with [\ion{O}{iii}] luminosity in a large sample of Seyfert galaxies
from Sloan Digital Sky Survey (SDSS, York et al 2000) Data Release 2
(DR2) (Simpson 2005; Hao et al. 2005). These results were
interpreted in the context of ``receding torus'' (Lawrence 1991;
Hill et al. 1996).  Similar results have been reported from
statistical studies of hard X-ray selected AGN (Ueda et al. 2003;
Hasinger et al. 2004; Gilli et al. 2007; Hasinger et al. 2008; c.f,
Wang et al. 2006).

Radio loud AGN, including radio quasars and radio galaxies, can also
be unified in such scheme (see Urry \& Padovani 1995 for a review).
As in the case for Seyfert galaxies, broad permitted lines were
detected in the polarized light of narrow line radio galaxies (e.g.,
Antonuuci, Hurt \& Kinney 1994; Tran, Cohen \& Goodrich 1995; Young
et al.1996; Tran et al. 1998). Barthel (1989) found that radio
galaxies have larger linear dimensions than that of radio quasars in
their 3CRR sample, and hypothesized that powerful radio galaxies and
radio quasars belong to the same population, with a cone angle of
$40^o -50^o$ to mark the division between the radio quasar and
galaxy. Within this scheme, by using 172 3CR radio sources, Lawrence
(1991) found that the type 2 fractions are anti-correlated with
radio luminosity in the range of $L_{\rm{178MHz}}\ge 10^{25}$ W
Hz$^{-1}$. Falcke et al. (1995) proposed that for radio-loud
sources, jets may clear a path through the dust, and cause the
obscuration along the jet's periphery, and give rise to the
anti-correlation between obscuration
 and radio luminosity. These mentioned above only focused on the most powerful
radio AGN.

In this paper, we will extend those analysis to radio
moderately-strong AGN, and examine in more details $f_1$ as a
function of radio power, [\ion{O}{iii}] luminosity, black hole mass
and accretion rate using the large sample of AGN in the SDSS
spectroscopic catalogues of galaxies and quasars that have been
detected in Faint Image of Radio Sources at Twenty cm ( FIRST, White
et al.1997). A combination of large sky coverage, high completeness
to a relatively deep magnitude for galaxies and quasars, and
moderate spectral resolution makes the sample ideal for such a
study.

Based on the SDSS spectroscopic data set, we culled 711 type 2
objects and 286 type 1 objects having FIRST luminosity
$P_{\rm{1.4GHz}} \geqslant 10^{23}$W Hz$^{-1}$ and $z\le 0.35$.
After correcting several selection effects, we obtained $f_1$ as a
function of nuclear parameters. The paper is arranged as follows.
The sample is described in the next section. The selection function
is estimated in \S 3 and the results are presented in \S 4. Finally,
we will discuss our result in \S 5. Throughout this paper, we will
adopt a concordant cosmology with $H_0 = 71$ km~s$^{-1}$~Mpc$^{-1}$,
$\Omega_{\rm {m}} = 0.27$, and $\Omega_{\rm{\lambda}}= 0.73$
(Spergel et al. 2003).

\section{The Sample of Radio Strong AGN}

\subsection{Parent Sample}

Starting with the spectroscopic samples of quasars and galaxies in
the SDSS data release four (DR4), we construct the low redshift
sample of radio detected galaxies and AGN. A redshift cut $z\leq
0.35$ is applied so that $\rm{H{\alpha}}$ falls in the SDSS spectral
coverage. To simplify the estimation of selection effects, we
consider only the objects targeted as main galaxies (Petrosian
magnitude $r\le 17.77$ and the target mask as ``TARGET\_GALAXY'') or
low redshift quasars (psf magnitude $i\le 19.1$ and the target mask
as ``TARGET\_QSOs\_CAP/SKIRT''), or FIRST counterparts
 (unresolved objects with FIRST counterparts within 2$\arcsec$ and psf magnitude $i\le
 19.0$, masked as ``TARGET\_QSOs\_FIRST\_CAP/SKIRT''). This low-$z$ galaxy and
quasar sample (405,904 SDSS spectra in total) are then
cross-correlated with the FIRST source catalog (Becker et al. 2003)
to form the radio detected galaxy and AGN sample with a procedure
described in Lu et al. (2007).

We use positional coincidence to select radio point sources, and
visually inspect the FIRST images for all the candidates with
extended radio morphology (refer to Lu et al. 2007 for details).
Briefly, we take 2$\farcs$0 as the cutoff of position offset for
SDSS-FIRST positional coincidence, which is a trade-off between the
completeness and random contamination. Then, we visually inspected
the cutouts of FIRST image to select the extended lobe(s) apart from
SDSS nucleus, using a degraded image of 3CR radio sources as the
reference for physical association. For $z \ge 0.073$, we checked
the FIRST cutouts in $6' \times6'$, which corresponds to a physical
size of 500 kpc at the redshift z=0.073, because only a small number
of radio sources have linear sizes larger than this. For $z \le
0.073$, we check the cutout with an angular size corresponding to
their linear sizes of 500 kpc at the sources' redshift. As the
outcome of the second stage, we obtain 20,334 low redshift
($z<0.35$) objects from SDSS-FIRST matching process.

Next, we apply a radio luminosity cutoff to the sample. According to
Yun et al. (2001) and Hopkins et al. (2003), star formation (SF)
galaxies rarely have radio powers larger than  $P_{\rm{1.4GHz}} =
10^{23}$W Hz$^{-1}$. Best et al. (2005) found that the radio
luminosity function at 1.4 GHz of SF galaxies intersect with radio
loud AGN at $\sim 10^{23}$W Hz$^{-1}$. Beyond this value, the
luminosity function of SF galaxies drops dramatically, while that of
radio AGN decreases mildly. Although at $\sim 10^{24}$W Hz$^{-1}$,
SF galaxies account for nearly 10\% of the population, they
 will be rejected on the emission line-ratio diagrams (hereafter, BPT diagram;
Baldwin, Phillips \& Terlevich 1981). So we will limit our analysis
to the sources with radio luminosities above $10^{23}$W Hz$^{-1}$.

A $k$-correction to the radio luminosity is applied by assuming a
radio spectral index $\alpha=0.5$ ($f_\nu\propto\nu^{-\alpha}$).
After retaining only the highest S/N spectrum for the duplicated
observations, we obtain 7,810 SDSS radio loud sources, 276 from SDSS
quasar sample and 7,534 from SDSS galaxy sample, at redshift $z \le
0.35$ and with $P_{\rm{1.4GHz}} \ge 10^{23}$W Hz$^{-1}$.

Our criteria are similar to Best et al. (2005), who cross-correlated
SDSS-FIRST ``compact'' sources with 3$\arcsec$ and selected the
``extended'' radio counterparts within 30$\arcsec$ by assuming that
radio sources are ``double lobe'' or ``core-lobe'', but extend to a
fainter limit and also include larger radio sources. These authors
extracted 2,215 radio AGN and 497 SF galaxies brighter than 5 mJy
(corresponding to $10^{23}$ W Hz$^{-1}$ at $z\sim$0.1) from 212,000
$z<0.3$ SDSS DR2 ``main galaxies'' spectroscopic targets, and their
AGN classification was based on the $D_n(4000)$ versus
$L_{\rm{1.4GHz}}/M_{*}$ diagnostic plane.

\subsection{Continuum Subtraction and Emission-Line Measurements}

In order to classify and obtain the intrinsic properties of the
radio galaxies and AGN, precise measurements of emission line fluxes
are necessary. We use the measured parameters from the Value-added
Extra-GAlactic Catalog developed and maintained by Center for
Astrophysics, University of Science and Technology of China
(USTC-VEGAC; X.-B. Dong et al. in preparation). We will describe
briefly the steps relevant here and leave details to the referred
paper. We correct the SDSS spectra for the Galactic extinction
(Schlegel, Finkbeiner \& Davis 1998) using the extinction curve of
Fitzpatrick (1999). The spectra are then brought into their rest
frame using the redshift provided by SDSS pipeline. The continuum
subtraction and emission line measurements are done separately for
the different type of objects as follows:

The continuum subtraction is done according to their relative
contribution of star-light and AGN continuum. For the narrow line
objects, the continuum is dominated by star-light, thus, can be
modeled with Independent Component templates (IC templates)
following the procedures described in Lu et al. (2006). \footnote{
IC templates were derived from Independent Component Analysis (ICA),
developed by Lu et al. (2006). Using this technique, Lu et al.
(2006) compressed the synthetic galaxy spectral library
  to six nonnegative independent components (ICs),
 which were proved to be good templates for modeling most normal galaxy spectra. }
 In brief, we fit galaxy spectra with the templates derived by applying Essembling
Learning for Independent Component Analysis (EL-ICA) to the simple
stellar population library (Bruzual \& Charlot 2003).
 The templates were then broadened and shifted to match
the stellar velocity dispersion of the galaxy. In this way, stellar
absorption lines are reasonably modeled to ensure the reliable
measurement of weak emission lines. At the same time, stellar
velocity dispersion, a correction to the redshift as well as an
average internal extinction to the stellar light is obtained.

For nucleus dominated type 1 AGN where \ion{Fe}{ii} multiplets and
other broad emission lines are highly blended, we fit simultaneously
the nuclear continuum, the \ion{Fe}{ii} multiplets and emission
lines (see Dong et al. 2008 for details). The nuclear continuum is
approximated by a broken power-law. \ion{Fe}{ii} emission, both
broad and narrow, is modeled using the \ion{Fe}{ii} templates
provided by Veron-Cetty et al. (2004). For those Seyfert 1s with
significant contribution of starlight as measured by the equivalent
widths (EWs) of the \ion{Ca}{ii k} $\lambda3934$ or high order
Balmer absorption lines or \ion{Na}{i} $\lambda\lambda5890,5896$, a
starlight model is also included using the 6 IC templates as
described above. The decomposition of host-galaxy starlight, nuclear
continuum and \ion{Fe}{ii} emission were carried out following the
procedure as described in detail in Zhou et al. (2006).

After subtracting the continuum, we fit emission lines with
multi-gaussian model using the code described in detail in Dong et
al. (2005, 2008). Briefly, each line is fitted with one or more
Gaussians as statistically justified (mostly with 1--2 Gaussians);
the line parameters are determined by minimizing $\chi^2$. The
[\ion{O}{iii}] $\lambda\lambda4959,5007$ doublet are assumed to have
the same profiles and redshifts; likewise, [\ion{N}{ii}]
$\lambda\lambda6548,6583$ and [\ion{S}{ii}]
$\lambda\lambda6716,6731$ doublet are constrained in the same way.
Furthermore, the flux ratios of [\ion{O}{iii}] doublet and
[\ion{N}{ii}] doublet are fixed to the theoretical values. Usually,
H$\alpha$ and [\ion{N}{ii}] doublets are highly blended and thus
hard to be isolated; in such cases, we fit them assuming they have
the same profile as [\ion{S}{ii}] doublets, which is empirically
justified (e.g., Filippenko \& Sargent 1988; Ho et al. 1997; Zhou et
al. 2006). For the possible broad H$\alpha$ and H$\beta$ lines, we
use multiple Gaussians to fit them, as many as they could be
statistically justified. If a broad emission line is detected with
S/N $> 5$, we regard it as genuine. If the broad H$\beta$ line is
too weak to achieve a reliable fit, we then re-fit it assuming that
it has the same profile and redshift as the broad H$\alpha$ line.

\begin{figure}
\begin{center}
\setlength{\unitlength}{1mm}
\begin{picture}(120,115)
\put(0,-35){\includegraphics{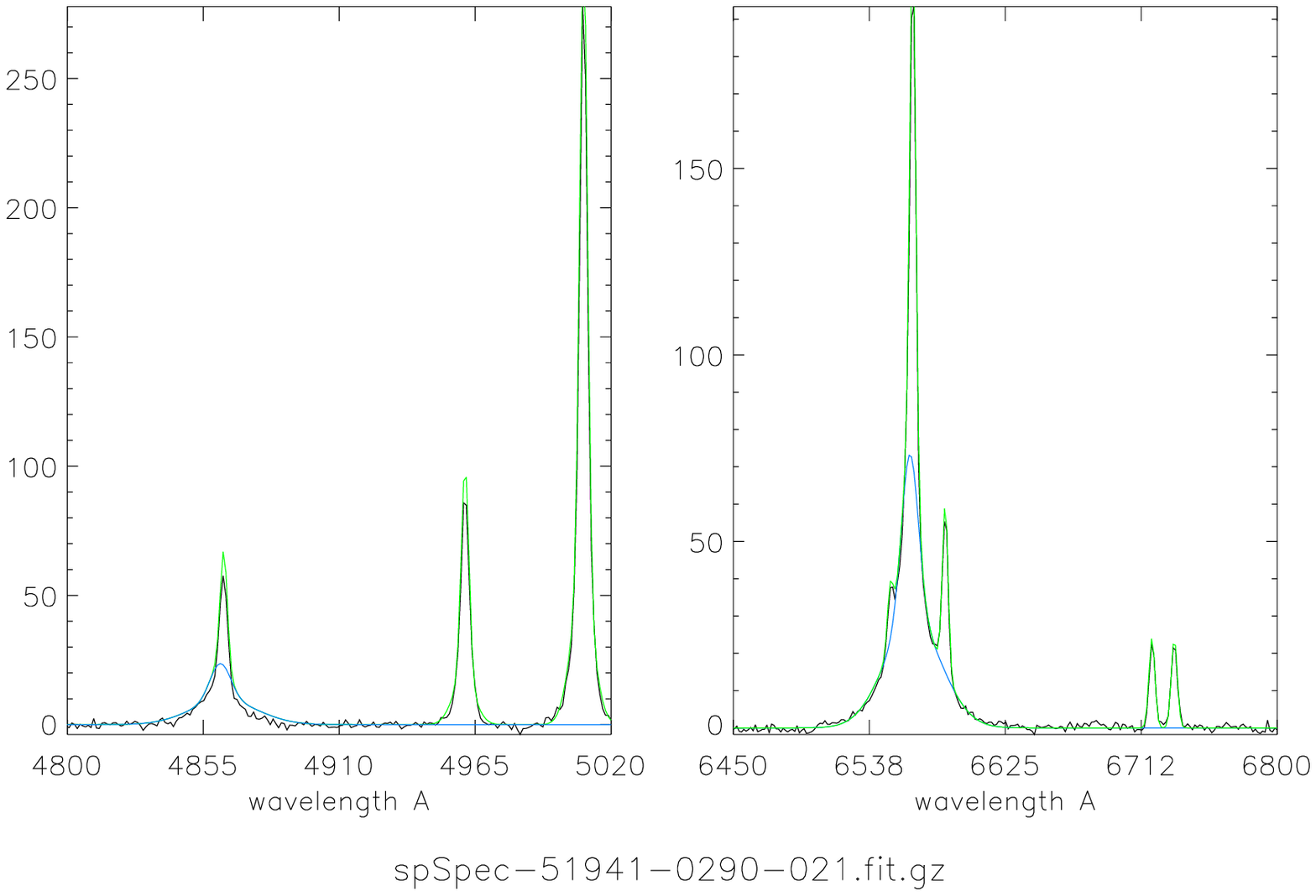}}
\put(0,-5){\includegraphics{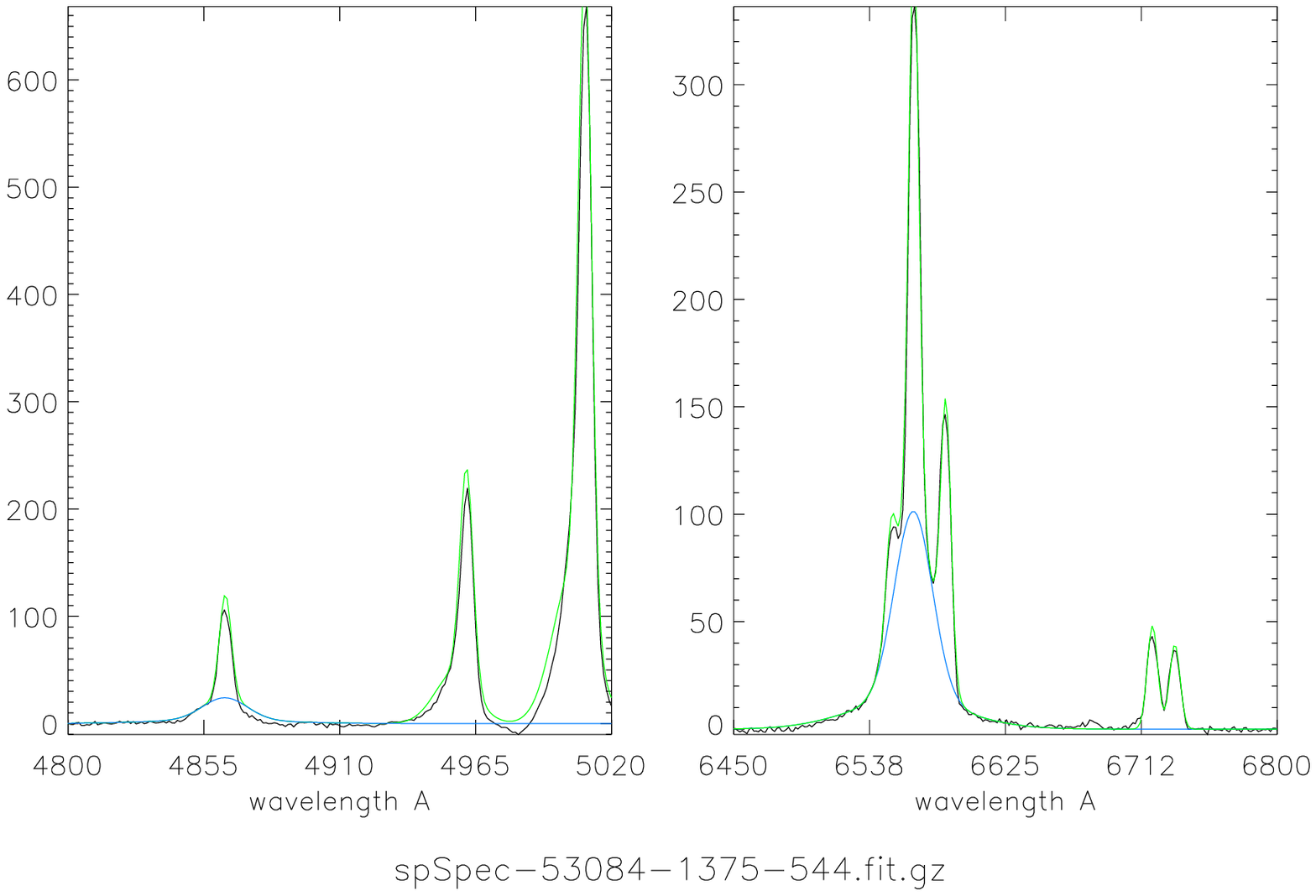}}
 \put(0,25){\includegraphics{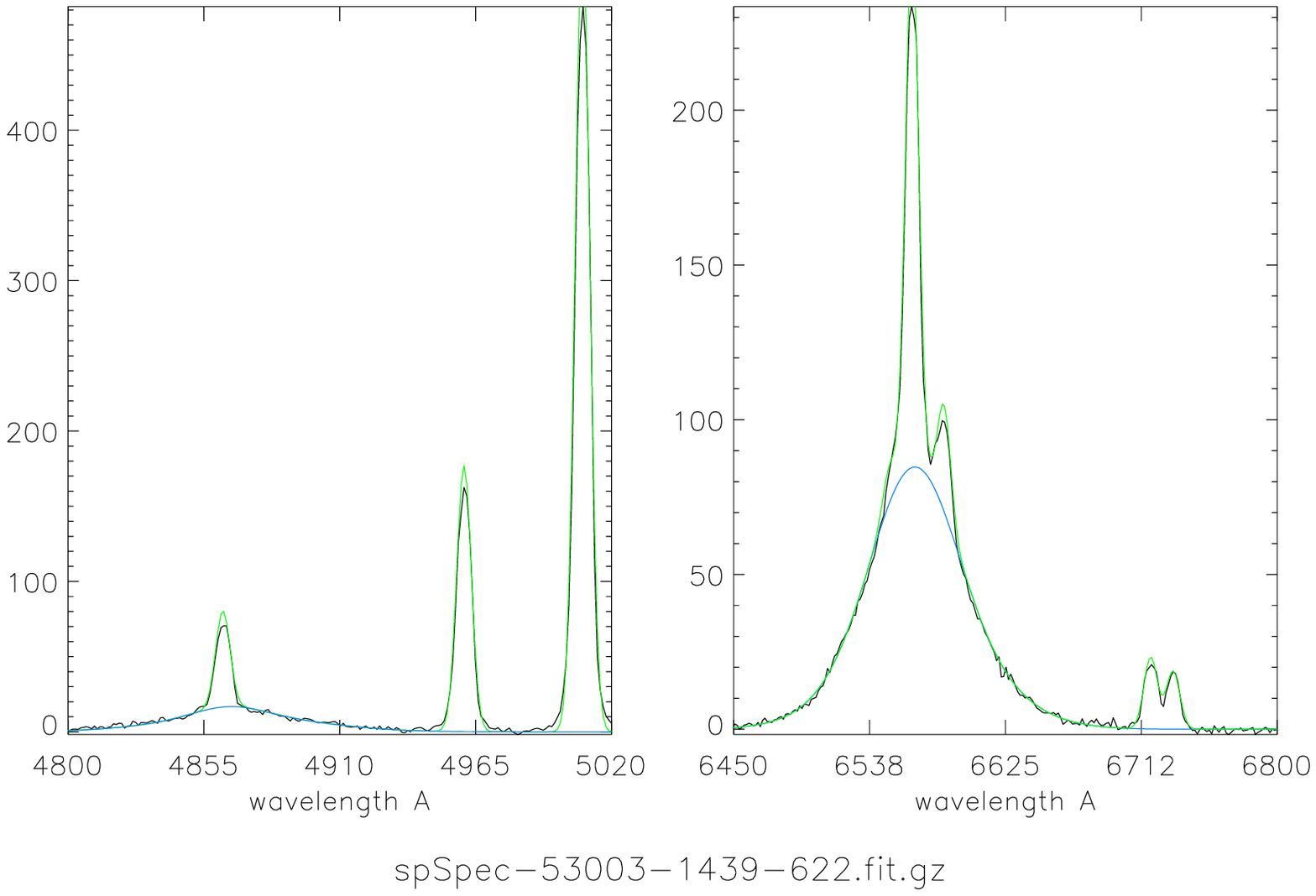}}
\put(0,55){\includegraphics{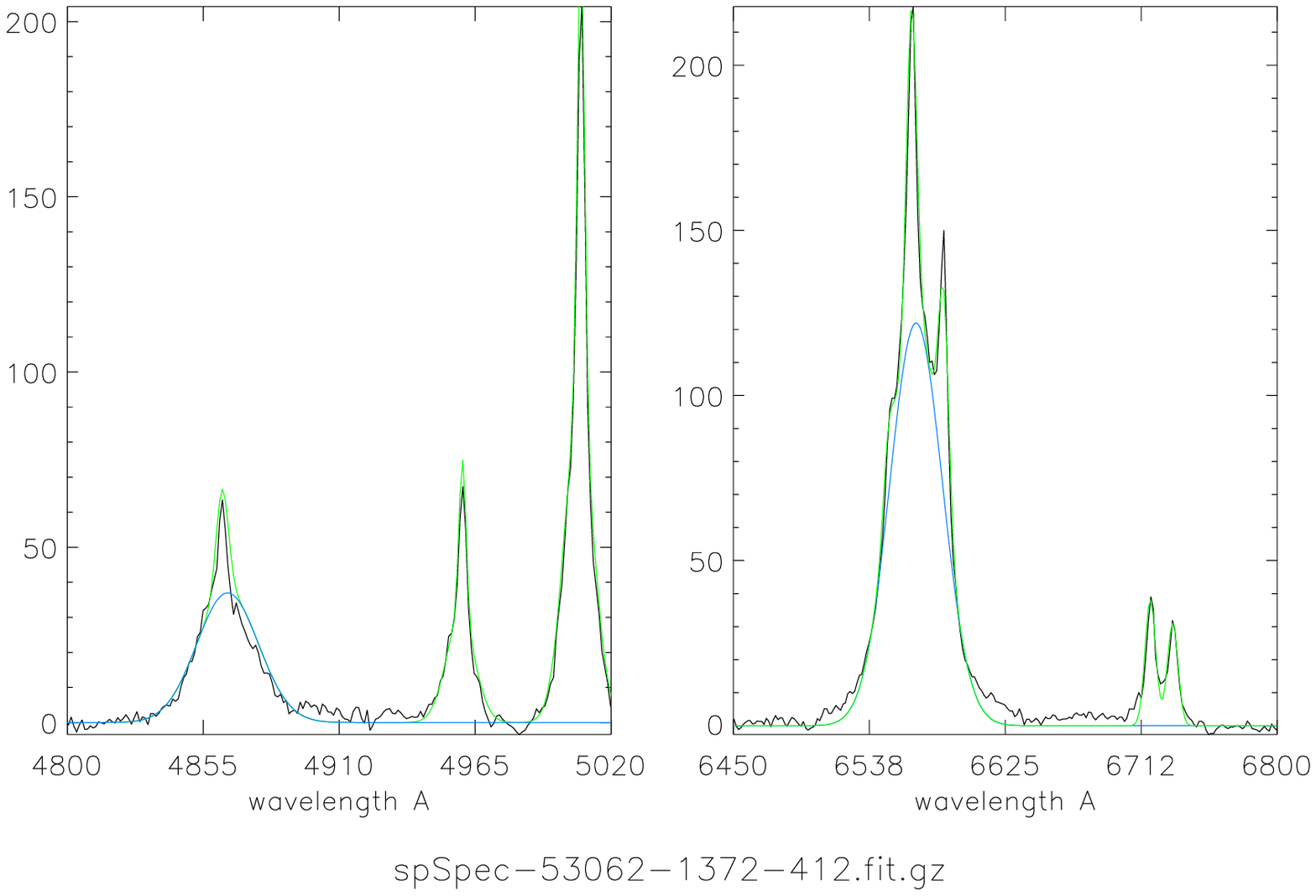}}
\end{picture}
\end{center}\caption[junk]{\label{fig:fitted_spectrums} Emission line fitting in the
 H$\alpha$ and H$\beta$ regions. The
black lines represent the observed ``continuum free'' emission lines
and noise. The blue lines represent for the fitted broad components
and the green lines for broad+narrow fitting. }
\end{figure}

Examples of emission line modeling in H$\alpha$+[\ion{N}{ii}] and
H$\beta$+[\ion{O}{iii}] regions are illustrated in Figure
\ref{fig:fitted_spectrums}. The two upper panels shown the objects
which was classified by SDSS as galaxies, and the two lower ones
shown the SDSS classified quasars. The total emission line flux is
estimated by adding different components, while the line width is
measured in the combined model.

\subsection{Optical Spectral Classification}

Based on the emission line parameters measured in the last section,
we classify optical spectra into broad lined and narrow lined AGN,
and SF galaxies, respectively. Detail criteria for each class are as
follows:

Broad line AGN are culled according to the following criteria (cf.
Zhang et al. 2008, X.-B. Dong et al. in preparation): (1) The
broad H$\alpha$ component has been detected with S/N $>5$;
 (2) The height of broad H$\alpha$ is more than twice of root mean
square (RMS) of the continuum-subtracted spectrum in the neighbor
emission-line free region to account for potential systematic errors
brought in the continuum subtraction; (3) The equivalent width of
broad H$\alpha$ line $EW($H$\alpha)\ge 10$\AA, in order to eliminate
potential contamination from the broad wing of the narrow component.
It should be mentioned that since we use strict criteria here to
select broad line AGN, we will correct the incompleteness late
through simulations.

\begin{figure}
\begin{center}
\setlength{\unitlength}{1mm}
\begin{picture}(150,28)
\put(0,-2){\includegraphics{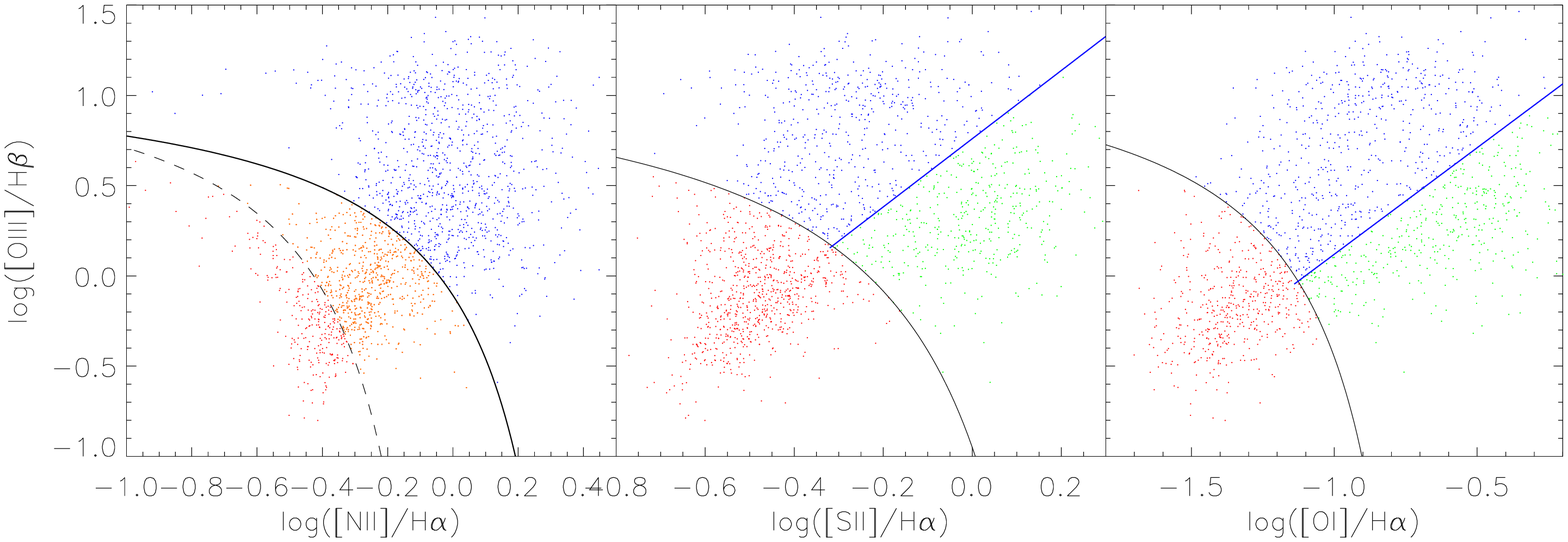}}
\end{picture}
\end{center}
{\caption[junk]{\label{fig:BPT} {The BPT diagram for narrow emission
line galaxies with radio luminosity $P_{\rm{1.4GHz}} \ge 10^{23}$W
Hz$^{-1}$ and $z\le0.35$. The solid lines separate Seyfert from SF
galaxies, and the dashed lines divide the ``composite'' galaxies
from SF galaxies. The blue line represents for Seyfert/LINER
separation.} }}
\end{figure}

Narrow line objects are classified into SF galaxies, Seyfert 2
galaxies (Sy2), Low Ionization Nuclear Regions (LINERs) and
composite galaxies, according to their locations on BPT diagrams. We
adopt the empirical classification scheme proposed by Kewley et al.
(2006). First, we divide narrow line radio galaxies into
star-forming galaxies, composite galaxies and 'pure' AGN on
[\ion{N}{ii}]/H$\alpha$ versus [\ion{O}{iii}]/H$\beta$ diagram.
Second, 'pure' AGN are further separated into 'LINER' or Seyfert
according to their locations on the [\ion{O}{iii}]/H$\beta$ versus
[\ion{S}{ii}]/H$\alpha$ or [\ion{O}{i}]/H$\alpha$ diagram (see
Figure \ref{fig:BPT}). For 137 objects with strong detections in
[\ion{O}{iii}] (with S/N$>10$) but none in H$\beta$ (S/N$<3$), we
estimate their H$\beta$ by using the average H$\alpha$/H$\beta$ of
$6.08$. Most non-detections are caused by low signal to noise of the
spectrum around H$\beta$ region or low H$\beta$ equivalent width,
but not due to large extinctions. Their 3 $\sigma$ upper limits are
consistent with the average value of H$\alpha$/H$\beta\sim6.08$.

With the above criteria, we obtain 352 broad lined objects. However,
among them, 43 objects (40 galaxies, 3 quasars) do not have reliable
broad H$\beta$ flux (with $\rm{S/N}<5$), 13 ones (11 galaxies, 2
quasars) have broad line Balmer decrement
(H$\alpha/$H$\beta)_{\rm{BL}}>10$, in analogy with type 1.8/1.9
Seyfert galaxies (cf. Dong et al. 2005). We reject those objects
from further analysis because: (1) the extinction correction to the
broad or narrow lines are uncertain for objects without H$\beta$
flux; (2) the sample becomes very incomplete at large Balmer
decrements due to their large obscuration.

Finally, 286 broad lined objects, including 248 from SDSS quasar
sample and 38 from the main galaxy sample, will be used for further
analysis. We note that 55 of these Type 1 AGN (46 quasars and 9
galaxies) are Narrow-line Seyfert 1 galaxies (NLS1's), according to
the criteria of Zhou et al. (2006), with broad component of
H$\alpha$ or H$\beta > 10$ $\sigma$ confidence level and
FWHM(H$\alpha) \le 2,200$ km s$^{-1}$. 35 of them are already
included in the NLS1 sample of Zhou et al. (2006).

\begin{figure}
\begin{center}
\setlength{\unitlength}{1mm}
\begin{picture}(150,30)
\put(0,-40){\includegraphics{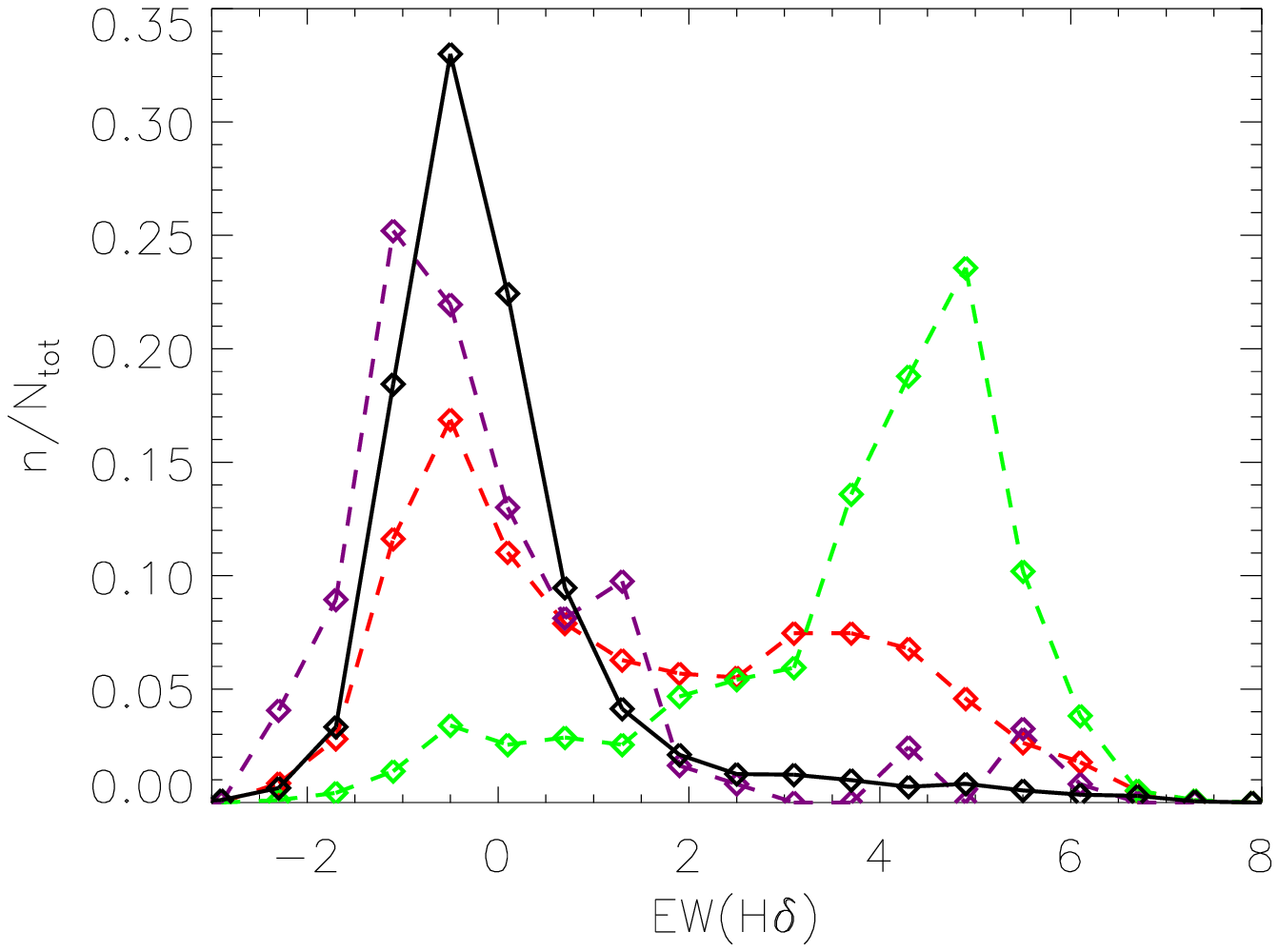}} \put(40,-40){\includegraphics{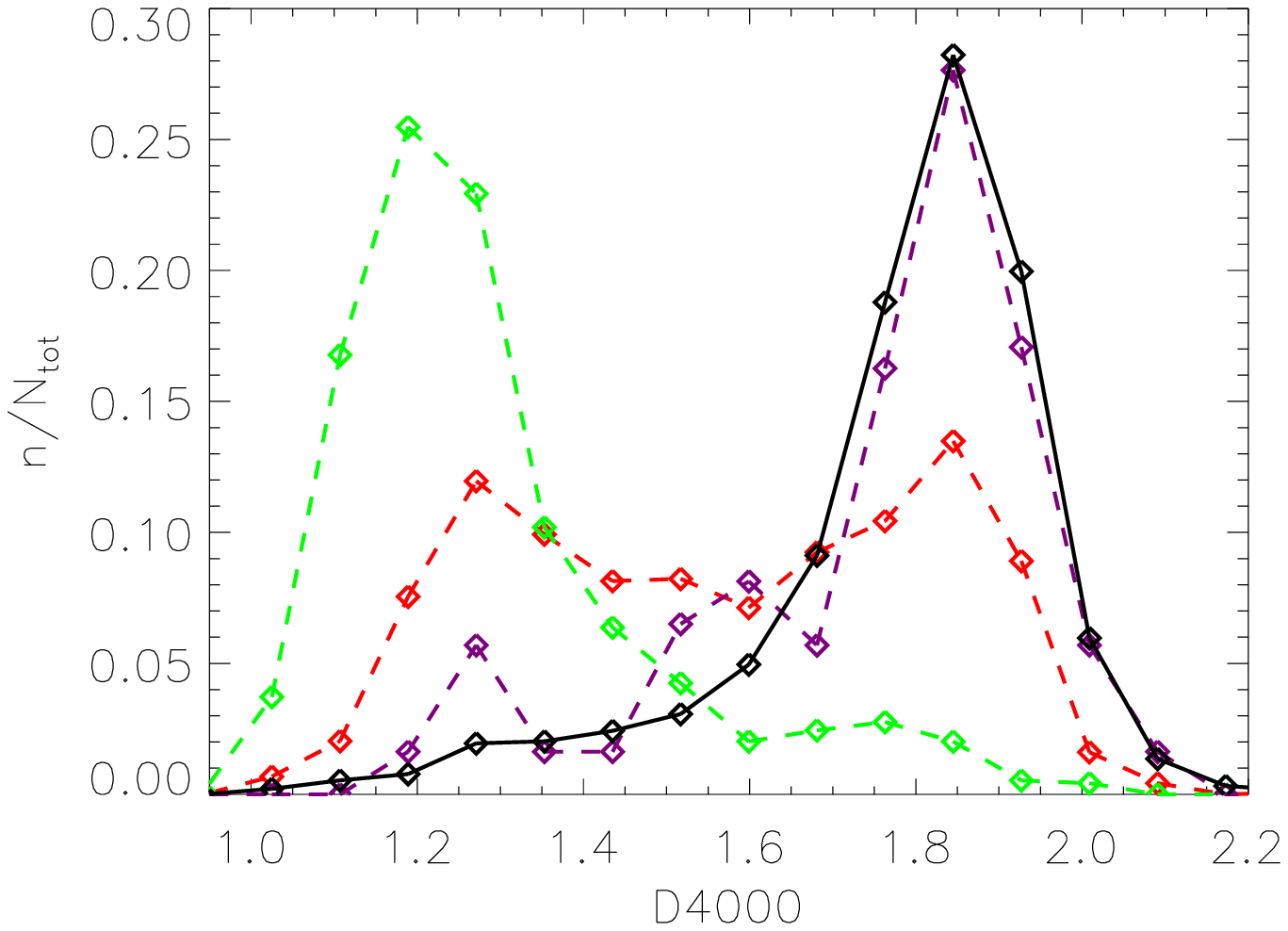}}
\end{picture}
\end{center}
{\caption[junk]{\label{fig:D4000_HdEW_test} {The distribution of
H$\delta$ absorption line equivalent width (left panel) and strength
of 4000\AA~ break (right panel) for Seyfert 2 galaxies (dashed red
line), SF galaxies (dashed green line), LINERs (dashed purple line),
and weak lined objects (solid black line).} }}
\end{figure}

Among the 3,297 galaxies with detectable emission lines, 1,771 ones
can be un-ambiguously classified, yielding 711 Seyfert 2 galaxies,
342 LINERS, 216 star-forming galaxies and 502 ``composite''
galaxies. The remaining 1,526 weak emission line galaxies either do
not have sufficient number of measured line ratios to allow a
meaningful diagnostic on the BPT diagram (1510 objects), or the
narrow line decrement H$\alpha$/H$\beta>15$ (16 objects). Most of
those weak lined objects are likely to be LINERs rather than Seyfert
2 galaxies because the distributions of their H$\delta$ absorption
line equivalent width and strength of 4000\AA~ break are more
similar to that of LINERs rather than Seyfert galaxies (see Figure
\ref{fig:D4000_HdEW_test}). Therefore, excluding those weak lined
objects will not seriously affect our estimate of type 1 fraction.

\begin{figure}
\begin{center}
\setlength{\unitlength}{1mm}
\begin{picture}(150,32)
\put(15,-6){\includegraphics{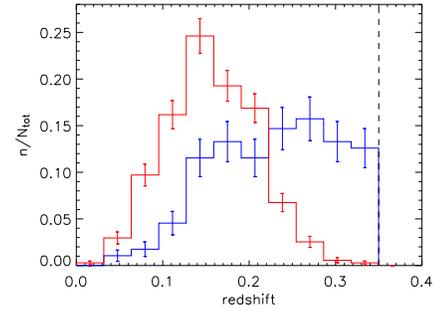}}
\end{picture}
\end{center}
{\caption[junk]{\label{fig:redshift} {The redshift distribution of
type 1s (blue line), and type 2s (red line).} }}
\end{figure}

\begin{figure}
\begin{center}
\setlength{\unitlength}{1mm}
\begin{picture}(150,32)
\put(15,-6){\includegraphics{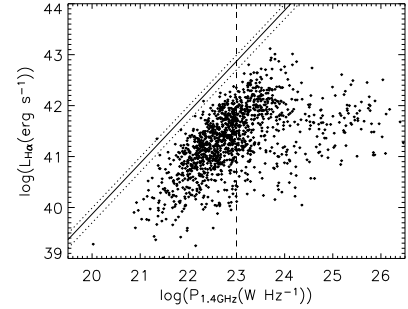}}
\end{picture}
\end{center}
{\caption[junk]{\label{fig:SFR} {Radio luminosity $P_{\rm{1.4GHz}}$
versus narrow H$\alpha$ luminosity $L_{\rm{H\alpha}}$. The points
represents for all SDSS selected radio AGN, regardless their radio
luminosity. The solid line indicates
$P_{\rm{1.4GHz}}-L_{\rm{H\alpha}}$ relation in SF galaxies. Beneath
this line, the radio luminosity $P_{\rm{1.4GHz}}$ will be barely
dominated by star formation. The dashed line indicates the $10^{23}$
W Hz$^{-1}$ threshold.} }}
\end{figure}

The final sample consists of 711 type 2 AGN and 286 type 1 AGN,
selected from original 7810 radio galaxies and quasars with $z \le
0.35$ and $P_{\rm{1.4GHz}}\ge 10^{23}$ W Hz$^{-1}$. Their redshift
distribution are shown in Figure \ref{fig:redshift}. Their locations
on the radio luminosity versus the narrow H$\alpha$ luminosity
$L_{\rm{H\alpha}}$ are shown on Figure \ref{fig:SFR}. For
comparison, we also add the empirical relation for SF galaxies (Yun
et al. 2001; Kewley et al. 2002). All radio AGN locate on the lower
right regime, suggesting that there is little radio contamination
from the star-forming process. This is true, even if the sample is
extended to a much lower radio power.

\subsection{Black Hole Mass, Nuclear Luminosity and Eddington Ratio}

For the Seyfert 2 galaxies, the black hole mass ($M_\bullet$) is
estimated from the stellar velocity dispersion using the
$M_\bullet-\rm{\sigma}_*$ relation obtained from local host
spheroids (Tremaine et al. 2002), including elliptical galaxies and
the bulges of disk galaxies. The intrinsic scatter of this relation
is estimated to be about 0.3 dex. The stellar velocity dispersion
has been obtained by fitting the SDSS spectrum with the IC templates
(refer to \S 2.2). It should be pointed out that the stellar light
within 3$\farcs$ of the SDSS fibre may contain only a fraction of
galaxy spheroid or have a substantial contribution from the galactic
disk. In the former case, the correction is usually small for most
of our galaxies according to the formula of Jorgensen et al. (1995).
In the latter case, the measured $\rm{\sigma}_*$ may over-estimate
the stellar velocity dispersion of the bulge component in high
inclination disk galaxies, or under-estimate the true value in low
inclination disk galaxies. The correction depends on the size of the
disk and bulge, the bulge disk ratio, the bulge velocity dispersion
to the circular velocity of the disk, the distance of the galaxy, as
well as the inclination of the disk. There is no simple formula for
this correction. Fortunately, most radio selected AGN have large
black hole masses, thus reside in the galaxies with large bulges. In
these galaxies, the disk-light contamination within SDSS fibre may
not be very severe. In addition, the rotational velocity is
proportional to the central stellar velocity dispersion of these
large galaxies (Courteau et al. 2008). Therefore, this correction
may be less severe to our sample.

\begin{figure}
\begin{center}
\setlength{\unitlength}{1mm}
\begin{picture}(140,35)
\put(25.5,-44.5){\includegraphics{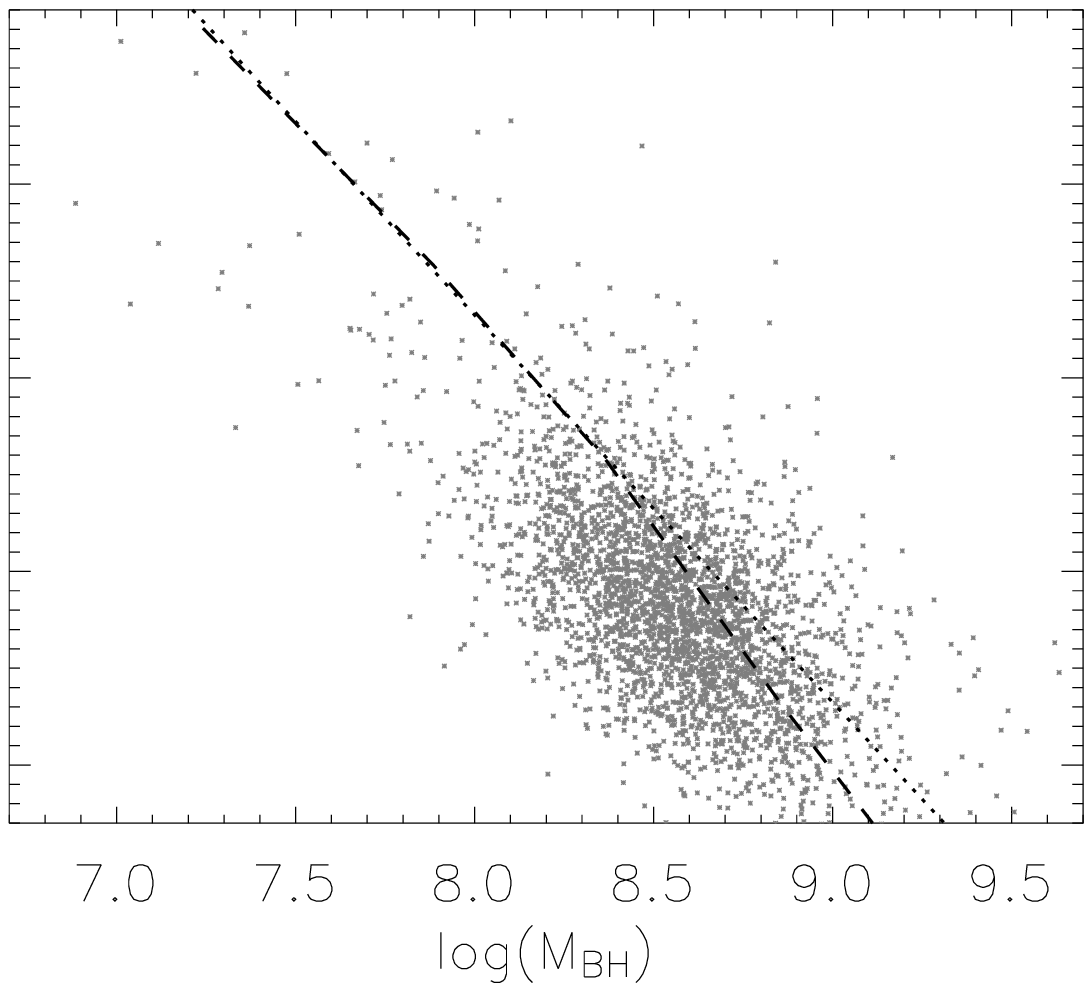}} \put(-6,0){\includegraphics{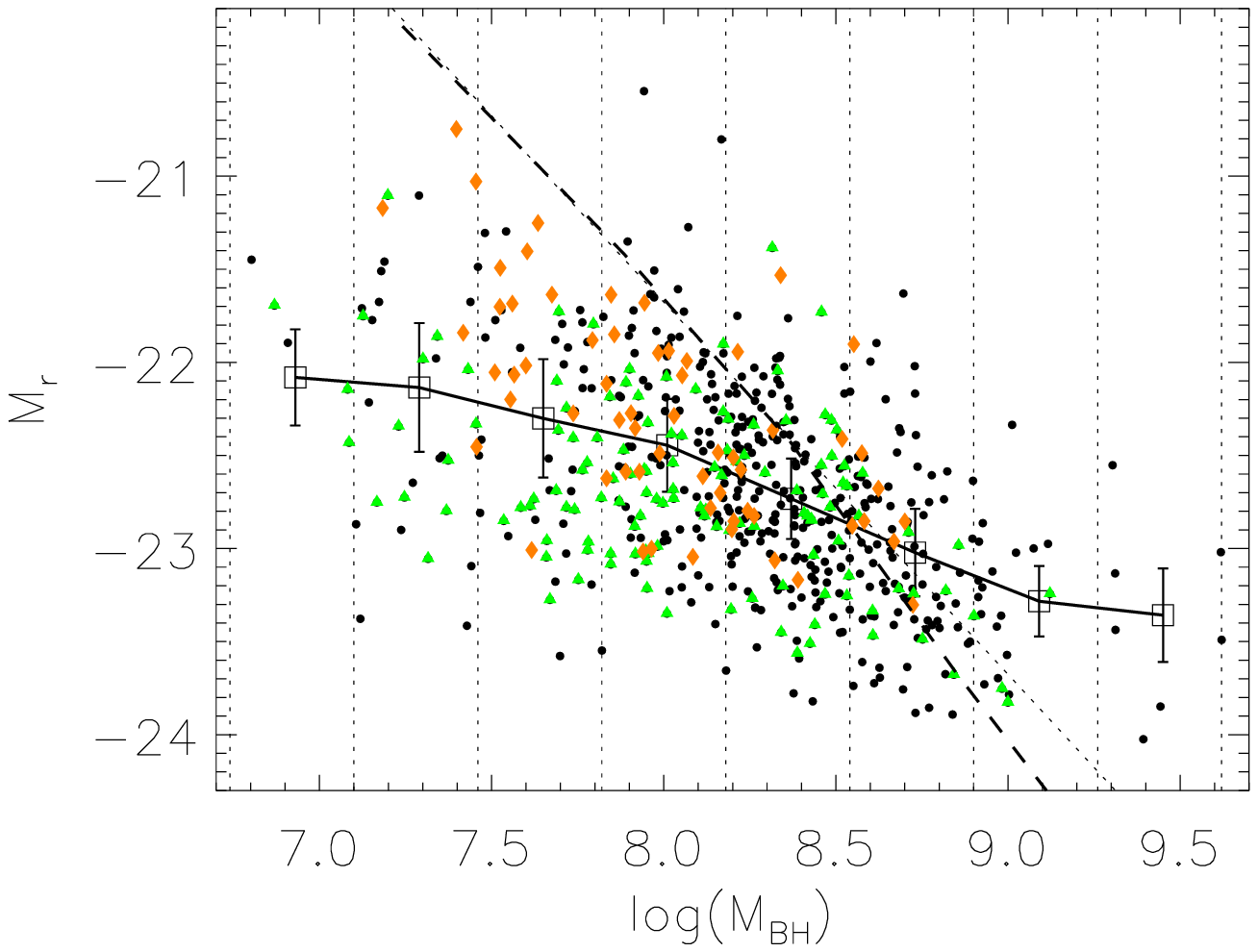}}
\end{picture}
\end{center} \caption{\label{fig:MBH_Mr}
Black hole mass $\log(M_\bullet)$ versus the absolute optical
magnitude $M_r$ for the radio galaxies. Left panel displays the
Seyfert 2 galaxies. The orange points represent for the high
inclination disk galaxies with axial ratio $b/a<0.5$ and
$\rm{\sigma}_*/err>10$, the green points for low the inclination
disk galaxies with $b/a>0.8$ and $\rm{\sigma}_*/err>10$, and black
points for elliptical galaxies. The solid line shows the mean $M_r$
value in the $\log(M_\bullet)$ bins. The dotted line show the
$M_r$-$M_\bullet$ fitting of inactive galaxies suggested by Mclure
\& Dunlop (2002), and the dashed line is referred from Faber-Jackson
relation of Desroches et al. (2007). The right panel displays the
$M_r$-$M_\bullet$ relation of the absorption line galaxies for
comparison. The entire objects in these two panels have radio
luminosity $\log(P_{\rm{1.4GHz}}\rm{/W\ Hz^{-1}})>23$.}
\end{figure}

In order to check whether the latter effect is important for our
sample, we divide the galaxies into the disk and bulge dominated by
using the profile parameters provided by SDSS pipeline. We consider
it a disk galaxy if the likelihood for the exponential profile fit
is larger than de Vacouleurs model fit by 0.2 dex, or inverse
concentration index $petroR50/petroR90>0.33$. Note that the
concentration index criterion $C_1 = 0.33$ from Shimasaku et al.
(2001) would induce a $15\sim20\%$ contamination from opposite types
of both disk galaxies and elliptical galaxies. But we consider this
contamination acceptable. We separate the disk galaxies into high
and low inclination groups according to their axial ratios (b/a;
$b/a<0.5$ for high inclination systems, and $b/a>0.8$ for low
inclination systems). Besides, all oblate ($b/a<0.3$) objects are
considered as high inclination disks, regardless of their radial
profile. Figure \ref{fig:MBH_Mr} shows their distribution on the
$M_r$ versus $M_\bullet$ diagram; where $M_r$ is the $k$-corrected
absolute Petrosian magnitude of the host galaxy in the galaxy rest
frame by interpolating the five SDSS apparent magnitudes with Spline
function. 
For comparison, we plot the Seyfert 2 galaxies of our sample in the
left panel and plot the absorption line dominated radio galaxies in
the right panel. We find: (1) absorption-line dominated radio
galaxies are concentrated on the high luminosity regime, while the
spectroscopically type 2 AGN spread more to the lower luminosities.
(2) $M_r$ vs $M_\bullet$ is tilted for type 2 AGN. It departs
significantly from what is defined by the absorption line galaxies
or the early type galaxies at low $M_\bullet$, and towards higher
luminosity. This can be attributed most likely to the disk light
contribution. (3) the high inclination galaxies (orange points) show
systematically lower luminosities than that of low inclination
galaxies (green points) at a given $M_\bullet$ (i.e.
$\rm{\sigma}_*$) despite of their overlap on the plot, and they
locate more closely to that of absorption line galaxies. Shao et al.
(2007) showed that 'edge-on' galaxies are on average 0.8-0.9 fainter
 in magnitude than that of 'face-on' galaxies in the $r$-band due
 to the dust extinction. This
seems sufficient to explain the offset between low inclination and
high inclination disk galaxies observed here without invoking any
additional effect of a disk component on the $\rm{\sigma}_*$.

For type 1 AGN, the virial black hole mass can be estimated using
the empirical relation between the size of BLR and continuum
luminosity, and the emission line width (e.g., Wandel et al. 1999;
Kaspi et al. 2005; Peterson \& Bentz 2006;). We use the broad
H$\alpha$ luminosity as a surrogate for the continuum luminosity
(Wang \& Zhang 2003), to account for the potential contamination
from the stellar-light in the weak BLR source and from nonthermal
jet emission in some cases, using the formula of Greene\& Ho (2007):
\begin{eqnarray}
\label{eqn:MBH_Type1}
 M_\bullet=(3.0^{+0.6}_{-0.5})\times
10^6\left(\frac{L_{\rm{H\alpha}}}{10^{42} \rm{erg
s^{-1}}}\right)^{0.45^+_-0.03}\\
\nonumber \times \left(\frac{FWHM_{\rm{H\alpha}}}{10^3\rm{km
s^{-1}}}\right)^{2.06^+_-0.06} M_{\sun}.
\end{eqnarray}
In section 3.3, this relation will be used also in the simulation of
selection effect for the potentially overlooked broad emission lines
in the observed type 2 AGN. The typical uncertainty in this relation
is estimated to be around 0.5 dex (Vestergaard \& Peterson 2006).

\begin{figure}
\begin{center}
\setlength{\unitlength}{1mm}
\begin{picture}(150,40)
\put(10,-5){\includegraphics{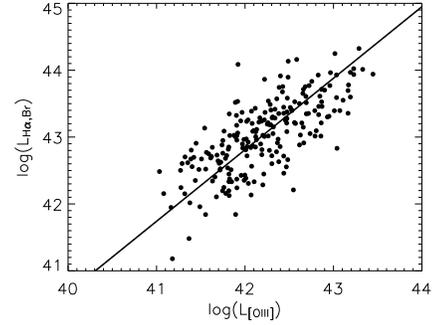}}
\end{picture}
\end{center}
{\caption[junk]{\label{fig:L_OIII-L_Ha} {The [\ion{O}{iii}]
luminosity $L_{\rm{[OIII]}}$ versus broad H$\alpha$ luminosity
$L_{\rm{H\alpha,Br}}$ . They are all extinction corrected on the
basis of their Balmer decrement H$\alpha$/H$\beta$.} }}
\end{figure}

We will use reddening corrected [\ion{O}{iii}] luminosity as an
indicator for the nuclear power of type 2 AGN (Mulchaey et al. 1994;
Dietrich et al. 2002; Kauffmann et al. 2003; Haas et al. 2007;
Netzer et al. 2006; See Reyes et al. 2008 for an extended
discussion) for two reasons. First, [\ion{O}{iii}] emission is less
contaminated by star-formation process, especially in these massive
 metal rich galaxies, than other lines such as H$\alpha$ and
[\ion{O}{ii}]. Second, Balmer decrement can be used as an indicator
for the global extinction to the narrow emission lines, thus the
intrinsic luminosity of [\ion{O}{iii}] can be recovered based on the
extinction corrected line flux.

The extinction correction to [\ion{O} {iii}] luminosity is still a
complex issue. Some authors argued that the some of the Balmer lines
may be totally blocked according to polarization observation (di
Serego Alighieri et al. 1997). Because the ionization potential of
O+ is much higher than that of hydrogen, [\ion{O} {iii}] emission
region may be even smaller than that of Balmer lines, and subject
more severely to the dust extinction. Therefore, extinction derived
from narrow line Balmer decrements may still underestimate the
[\ion{O}{iii}] extinction. However, most of
 the former arguments are based on the directly observed [\ion{O}{iii}]
luminosity rather than on the extinction corrected [\ion{O}{iii}]
luminosity. We consider applying correction should be better than
applying no correction.

We calibrate the relation between the bolometric luminosity and
[\ion{O}{iii}] luminosity with the broad line radio AGN and then
applied it to the type 2 objects as follows. First, we establish a
relation between the extinction corrected [\ion{O}{iii}] luminosity
and the broad H$\alpha$ luminosity for the type 1 radio AGN. The
extinctions are estimated based on the Balmer decrements of narrow
and broad lines, respectively, assuming an intrinsic
$\rm{H\alpha/H\beta}=3.1$. [\ion{O}{iii}] and broad H$\alpha$
luminosities are fairly well correlated (see Figure
\ref{fig:L_OIII-L_Ha}).

\begin{figure}
\begin{center}
\setlength{\unitlength}{1mm}
\begin{picture}(150,40)
\put(8,-6){\includegraphics{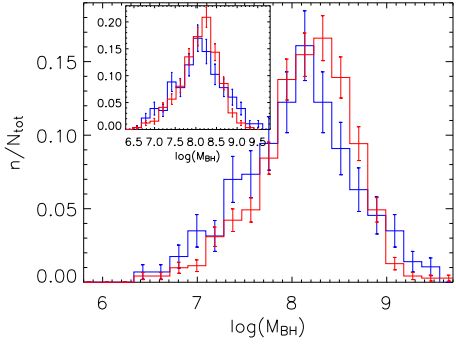}}
\end{picture}
\end{center}
{\caption[junk]{\label{fig:MBH} {The black hole mass $M_\bullet$
distribution of type 1s (estimated from reverberation-mapping, blue
line) and that of type 2s (estimated from bulge-$M_\bullet$
correlation, red line). The inset small diagram shows $M_\bullet$
distribution within restricted $L/L_{\rm{Edd}}$
   and $L$ regime, refer to Figure \ref{fig:selection_func} and \S 4.2.} }}
\end{figure}

\begin{figure}
\begin{center}
\setlength{\unitlength}{1mm}
\begin{picture}(150,40)
\put(8,-6){\includegraphics{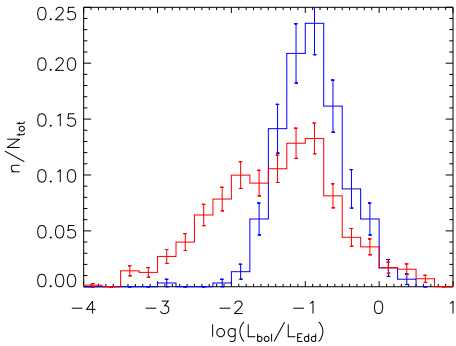}}
\end{picture}
\end{center}
{\caption[junk]{\label{fig:Eddington_ratio} {Eddington ratio
$\log(\ell)=\log(L/L_{\rm{Edd}})$ distribution for type 1s (blue
line) and type 2s (red line). } }}
\end{figure}

\begin{figure}
\begin{center}
\setlength{\unitlength}{1mm}
\begin{picture}(150,40)
\put(15,-6){\includegraphics{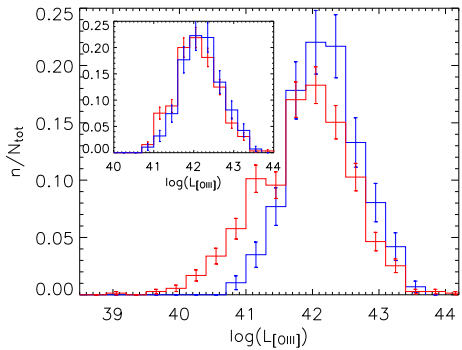}}
\end{picture}
\end{center}
{\caption[junk]{\label{fig:L_OIII} {The extinction corrected
[\ion{O}{iii}] luminosity distribution for type 1s (blue line) and
type 2s (red line). The small diagram shows $L_{\rm{[OIII]}}$
distribution within restricted $\log(L/L_{\rm{Edd}})$ and
$M_\bullet$ regime, refer to Figure \ref{fig:selection_func} and \S
4.1.} }}
\end{figure}

A least-square logarithmic-linear fit yields
\begin{equation}
\label{eqn:LOIII_LHa} \log(L_{\rm{H\alpha,
Br}})=(1.422\pm0.069)+(0.984\pm0.067)\times\log(L_{\rm{[O III]}})
\end{equation}
, with a scatter in $\log(L_{\rm{H\alpha, Br}})$ of 0.39. Note that
the slope is close to one, suggesting a linear relation between the
two luminosities, which is consistent with Zhang et al. (2008) for
radio quiet AGN.

Second, we estimate the continuum luminosity at 5100\AA~ from the
broad H$\alpha$ luminosity using the relation obtained by Greene \&
Ho (2007), for the nuclear light dominated AGN. Finally, bolometric
luminosity is estimated as  $L=9(\lambda L_\lambda)_{5100\AA}$. With
the bolometric luminosity and the black hole mass, it is straight
forward to calculate the Eddington ratio
$\ell=L_{\rm{bol}}/L_{\rm{Edd}}$.

The distributions of the black hole mass for the type 1 and type 2
AGN are displayed in Figure \ref{fig:MBH}. On average, the type 1
AGN have slightly smaller black hole mass than that of type 2 AGN.
The medians are 1.0$\times10^8~M_{\sun}$ and
1.4$\times10^8~M_{\sun}$ for the type 1 and type 2 AGN,
respectively. However, the type 1 AGN are on average 1.4 times more
luminous than type 2 AGN in the extinction corrected [\ion{O}{iii}]
luminosity (Figure \ref{fig:L_OIII}). Jackson \& Browne (1990) found
that radio quasars are a factor 10 more luminous in the
[\ion{O}{iii}] line luminosities than that of radio galaxies, and
they interpreted their results as a larger extinction to narrow line
region in radio galaxies than that in quasars. The difference found
here is much smaller because we have corrected the intrinsic
reddening using Balmer decrement while Jackson \& Browne have not.
Indeed, the average [\ion{O}{iii}] luminosity of the type 1 AGN is
more luminous by a factor of 6 than type 2 AGN before the extinction
correction. Heckman et al. (1992; see also Meisenheimer 2001) found
that the mid- to far-IR emission ($\lambda\sim 6-50$ $\mu m$ in AGN
rest frame) is 4 times stronger on average in quasars than that in
NLRG for a 178MHz 3CR sample (including 42 quasars and 75 NLRG with
$z>0.3$). If the mid-to-far-infrared emissions are isotropic and
powered by AGN, this implies that some of the NLRG may not be
Seyfert 2 type but be low luminosity LINERs, or the transition of
the broad line strength occurs in the low Eddington ratio and low
luminosity region (see \S 5).

The Eddington ratio $\log(\ell)=\log(L_{\rm{bol}}/L_{\rm{Edd}})$
distribution was displayed in figure \ref{fig:Eddington_ratio}. We
found that the distribution of the type 1 AGN peaked at
$\log(\ell)\sim-0.9 (\ell\sim0.13)$ in the range of $-2 \leq
\log(\ell) \leq 0$, while the distribution of the type 2 objects is
wider and skewed to a lower $\log(\ell)$ although it is also peaked
near $\log(\ell)\sim -0.9$. The more extended tail towards small
$\log(\ell)$ in the type 2 AGN may be attributed to a selection
effect because the broad line components are more difficult to be
detected at a lower $\log(\ell)$. Note that scatters in the
estimation of black hole mass and bolometric luminosity will cause a
distortion in $\log(\ell)$ distribution. Fortunately, the scatter
(0.5 dex) of the black hole estimation for type 1 using the line
width and continuum luminosity is comparable to the combination of
the scatter (0.3 dex) in $M_\bullet-\sigma_*$ relation and the
scatter (0.4 dex) in the bolometric luminosity estimation using
[\ion{O}{iii}] luminosity for type 2. Thus, the scatter effect on
type 1 and type 2 distributions is similar. We will discuss this
further in Section 4.

\section{Selection Effects}

Selection effects are introduced during the spectroscopic target
selection and the definition of narrow and broad line AGN sample. In
this section we will quantify these selection effects as a function
of nuclear properties for both the type 1 and type 2 AGN. We denote
the number density of AGN in unit nuclear luminosity ($L$) and unit
$M_\bullet$ intervals as $\phi(L, M_\bullet)$. The expected number
of AGNs in the nuclear luminosity bin $L-L+\Delta L$ and black hole
bin $M_\bullet-M_\bullet+\Delta M_\bullet$ can be written as,
\begin{eqnarray}
\label{eqn:selected_number} \Delta N(L, M_\bullet)=\phi(L,
M_\bullet) \int_{L^{s}_{min}(z)} p(L^{s}|M_\bullet,L)
S(M_\bullet,L,L^{s}) \\
\nonumber \times dL^{s} V_{\rm{max}}(L^{s}) \Delta M_\bullet \Delta
L
\end{eqnarray}
where $L^{s}$ is the luminosity of the band that is used in defining
the magnitude limit of the sample, e.g., the $r$-band luminosity of
the host galaxy dominated targets and $i$-band luminosity for the
nuclear dominated targets. $p(L^{s}|M_\bullet,L)$ is the conditional
probability that an AGN has $L^{s}$ at the given $M_\bullet$ and
$L$. Selection function $S(M_\bullet,L,L^{s})$ is the probability to
classify the spectrum as type 1 or type 2; $V_{\rm{max}}$ is the
comoving volume corresponding to the maximum redshift that an object
with a luminosity $L^{s}$ could be detected within the magnitude
limit. We have assumed that there is no cosmological evolution for
both types of AGN within $z<0.35$, and we will check the assumption
in \S 3.1 and \S 4.1. Since we are interested in the ratio of the
two type objects, and the sky coverage for type 1 and type 2 are the
same, we will not consider the sky coverage for the sample in the
volume calculation.

Rewriting Eq \ref{eqn:selected_number} for the binned data set, we
find
\begin{equation}
\phi(L, M_\bullet)=\sum_{i=1}^N \frac{1}{V_{max,i}}
\frac{h(L,L+\Delta L;M_\bullet,M_\bullet+\Delta
M_\bullet)}{\int_{L^{s}_{lim}}^{\infty}{p(L^{s}|M_\bullet,L)
S(M_\bullet,L,L^{s}) dL^{s}}} \label{eq:phi_L_M}
\end{equation}
for $N$ AGN in the sample, and $h(L,L+\Delta
L;M_\bullet,M_\bullet+\Delta M_\bullet)$ is the rectangular
function. Due to the additive nature of Eq \ref{eq:phi_L_M}, AGN
selected via exclusive rules can be added up simply.

With $\phi(L, M_\bullet)$, we can obtain the density of AGN in the
unit interval of Eddington ratio $\ell=L_{\rm{bol}}/L_{\rm{Edd}}$
and $M_\bullet$:
\begin{equation}
\label{eq:phi_M_lambda}
 \varphi(\ell, M_\bullet)= \int\phi(L,M_\bullet)\delta
(\ell-\ell(L))(\partial\ell(L)/\partial L)_{M_{\bullet}}dL
\end{equation}
The density at radio luminosity bin could be wrote as
\begin{equation}
\label{eq:phi_Lradio}
\psi(P_{\rm{1.4GHz}})=\int\phi(L,M_\bullet) p(P_{\rm{1.4GHz}}|L,M_\bullet) dL dM_{\bullet}
\end{equation}
, where $p(P_{\rm{1.4GHz}}|L,M_\bullet)$ is the conditional
probability that an AGN has a radio luminosity of $P_{\rm{1.4GHz}}$ at a
given $M_\bullet$ and $L$.

\subsection{ $V_{\rm{max}}$}

Different magnitude limits have been used in the selection of
different type spectroscopic targets. The apparent magnitude limits
for main quasars and FIRST counterparts are $i_{\rm{psf}} \leq 19.1$
on the psf magnitude, while for galaxies it is $r_{\rm{Petrosian}}
\leq 17.77$ in $r$ band. We calculate $V^{\rm{opt}}_{\rm{max}}$ for
each object according to its relevant optical magnitude limit and
the corresponding optical luminosity\footnote{Because we are
interested in the ratio of type 1 and type 2 AGNs, rather than their
comoving density, we do not take the survey area into
consideration.}. Since our objects were detected by FIRST, which has
a detection limit of $f_{\rm{1.4GHz}}\ge 1 $mJy, a
$V^{\rm{radio}}_{\rm{max}}$ can be estimated from the radio power
and the radio flux limit for each object. Furthermore, we adopt a
cutoff in the redshift $z \leq 0.35$, so the maximum volume is
$V_{0.35}$. The $V_{\rm{max}}$ is the smallest one among the above
three values.

For objects uniformly distributed in the unverse, the average
$<V/V_{\rm{max}}>$ will be around 0.5. We calculated this value for
both type 1 and type 2 AGN. We find $<V/V_{\rm{max}}>=0.55$ for the
type 1 AGN, and $<V/V_{\rm{max}}>=0.54$ for the type 2 AGN. This may
be taken as an evidence that the number density of AGN increases
mildly with increase redshift, and we will discuss the evolution
effect in \S 4.1.

\subsection{The Probability Function for Type 2s}

Ignoring the contribution of emission lines to the optical
magnitude, the type 2 AGNs are selected according to their host
galaxies. In order to quantify the conditional probability
$p(L^{s}|M_\bullet,L)$ of type 2 AGN for a given black hole mass and
nuclear luminosity (or [\ion{O}{iii}] luminosity), we need to
establish a relation among the host galaxy magnitude $L^s$, the
[\ion{O}{iii}] luminosity $L_{\rm{[OIII]}}$, and the black hole mass
$M_\bullet$. The strong correlation between the mass of the massive
black hole and the bulge luminosity of its host galaxy was
established for the bulge dominated quiescent galaxies in the local
universe (e.g., H{\"a}ring \& Rix 2004) and for the active galaxies
and quasars (e.g., Peng et al. 2006). However, there is still
controversial whether this relation depends on the level of nuclear
activity (McLure \& Dunlop 2002). To check this, we divide the type
2 AGN into high ($\log(L_{\rm{[OIII]}}\rm{/erg\ s^{-1}})>42$) and
low ($\log(L_{\rm{[OIII]}}\rm{/erg\ s^{-1}})<42$) [\ion{O}{iii}]
luminosity groups, and examine their distributions on
 $M_r-M_\bullet$ diagram. We find that at the same $M_\bullet$,
 the [\ion{O}{iii}] luminous galaxies are brighter than the low
$L_{\rm{[OIII]}}$ counterparts for only 0.1-0.2 mag on average,
which can be accounted for by their different emission line
luminosities. Therefore, we will assume that the host's luminosity
is independence of the nuclear luminosity at a giving black hole
mass. Therefore, we write $p(L^{s}|M_\bullet,L)$ as
$p(L^{s}|M_\bullet)$, i.e., independence of the nuclear luminosity
$L$.

The galaxies are selected based on their Petrosian magnitudes, i.e.,
$L^s$ should be the total magnitude of the galaxy. In order to
quantify $p(L^{s}|M_\bullet)$, we need to characterize the relation
between the black hole mass and the total luminosity of its host
galaxy, and the scatter of this relation as well. For the known
Seyfert 2 galaxies, their distributions are already shown in Figure
\ref{fig:MBH_Mr}. We assume the including of the missed Seyfert
galaxies does not change this relation. Because the disk
contribution increases as the black hole mass decreases, the
relation between the galaxy luminosity and black hole mass is
flatter than the bulge-black hole relation shown in Figure
\ref{fig:MBH_Mr}. For comparison, we also show the
$M_\bullet-M(\rm{bulge})$ relation for the inactive galaxies: $\log
(M_\bullet)=-0.50(\pm 0.05) M_R(\rm{bulge})-2.91(\pm 1.04)$ (McLure
\& Dunlop 2002), and the Faber-Jackson relation of Desroches et al.
(2007) for the normal elliptical galaxies. Apparently, $M_r- \log
(M_\bullet)$ relation of our sample is much flatter than the latter
relations. To describe quantitatively the distribution of the $M_r$
as a function of $M_\bullet$, we calculate the average value and the
second momentum of $M_r$ over each $M_\bullet$ bin, and
approximately estimate the $p(L^s|M_\bullet)$ with a gaussian
distribution around their mean value.

We assume the type 2 galaxies are relative easy to identify, so
$S(M_\bullet, L, L^s)=1$ once the object is observed
spectroscopically. This assumption will fail in two cases: (1)
nuclear emission lines are too weak, so that some lines used in the
spectral classification cannot be detected in the spectrum with a
typical signal to noise ratio of the main galaxy sample; (2) the
contamination of the emission lines from the HII regions of its host
galaxy makes the spectral classification as composite or HII type.
In the first case, we will miss very weak type 2 AGN, i.e., the low
Eddington ratio objects. But this will not affect our study of the
type 1 to type 2 ratios, because the detection of a type 1 AGN will
require an even higher Eddington ratio, and our analysis is limited
 to the parameter space that a substantial fraction of both the type 1
and type 2 can be detected (see Figure \ref{fig:selection_func}). In
the second case, we will drop some of the type 2 AGN in the late
type of galaxies, which tend to have lower black hole masses. We
will discuss this in \S 4.2.

\subsection{The Selection Function for Type 1s}

The probability of identifying a type 1 object depends on the signal
to noise ratio of the spectrum and broad line parameters, its
profile and intensity, in a rather complicated manner. To quantify
such a selection effect, we generate a large number of spectra
covering the H$\alpha$ blending and H$\beta$ regimes using
Monte-Carlo simulations similar to what has been done by Hao et al.
(2005) but taking additional parameters, like the intrinsic
reddening and black hole mass, into consideration. These spectra are
modeled in exactly the same way as we did for the real data to
obtain the emission line parameters, and the type 1 objects are
selected with the same criteria as described in \S 2.3. This will
give the probability of selecting a type 1 AGN under different
physical parameter regime.

A simulated emission line spectrum is the sum of three components:
the narrow line spectrum ($f_{narrow}$), broad line spectrum
($f_{broad}$) and the noise spectrum ($noise$). In order to mimic
the diverse narrow line spectrum and to avoid complicated noise
model, we use the observed narrow line spectrum and noise spectrum.
In order to explore the physical parameters as broad as possible, we
take the narrow line plus noise spectrum from both the type 1 and
type 2 AGN.

The narrow line plus the noise spectrum ($f_{narrow}+noise$) is
obtained by subtracting the power-law continuum and FeII models for
the nuclei dominated objects, or by subtracting the stellar
continuum model for the host galaxy dominated objects from the
observed spectrum. It should be noted that this treatment of noise
spectrum is in-exact for the Seyfert 2 galaxies, because the
addition of a broad line component and its corresponding nuclear
continuum would also increase the noise. But its effect is likely to
be small because a broad line with a height of ten percent of the
continuum flux would be easily detectable in most spectra.

For the broad line spectrum ($f_{broad}$), we estimated them with
 the parameters of line profile and line flux, which in turn depends on
the intrinsic broad-line luminosity and the dust extinction
to the BLR. We denotes the broad line component as,
\begin{equation}\label{eqn:simu_f_broad}
f_{broad}=A(\lambda) K f^{s}_{broad}
\end{equation}
where $A(\lambda)$ and $K$ are the internal extinction and the
scaling factor, respectively, and $f^{s}_{broad}$ is 'standard'
broad line spectrum.

For the type 1 objects, $f^s_{broad}$ is the best fitted model for
both broad H$\beta$ and H$\alpha$ after correcting for the internal
reddening. For the Seyfert 2 objects, we assume that the broad line
profile can be approximated with a single gaussian, and the line
width can be estimated from the empirical relation in
Eq.\ref{eqn:MBH_Type1}, with the black hole mass estimated from
$M-\rm{\sigma}_*$ relation, and line-flux obtained from the observed
[\ion{O}{iii}] luminosity with Eq.\ref{eqn:LOIII_LHa}.

$A(\lambda)$ comes from a set of eight extinction values, which
corresponds to a uniformly distributed H$\alpha$/H$\beta$ values
between 3-10, assuming the intrinsic H$\alpha$/H$\beta$=3.1. For a
given $A(\lambda)$, a set of ten $K_n$ are created randomly
following the H$\alpha$ luminosity distribution at the observed
[\ion{O}{iii}] luminosity, i.e.,
$p(L_{\rm{H\alpha}}|L_{\rm{[OIII]}})$ (see Figure
\ref{fig:L_OIII-L_Ha}).

To summarize, for each AGN, we build a set of simulated spectra that
has identical narrow line+noise spectrum as the observed spectrum
but with a series of manually-built broad line components. Assuming
the nuclei luminosity scaled with broad line, the apparent magnitude
and volume limit will change when we vary $f_{broad}$ component of
Seyfert galaxies. We retained only those $K_n$ and $A(\lambda)$ that
makes the optical flux for the simulated spectrum within the
magnitude limit ($r<17.77$ for the host dominated objects and
$i<19.1$ for the nuclear dominated objects). As a result, up to 80
spectra are created for each observed AGN.

\begin{figure}
\begin{center}
\setlength{\unitlength}{1mm}
\begin{picture}(150,33)
\put(-4,-2){\includegraphics{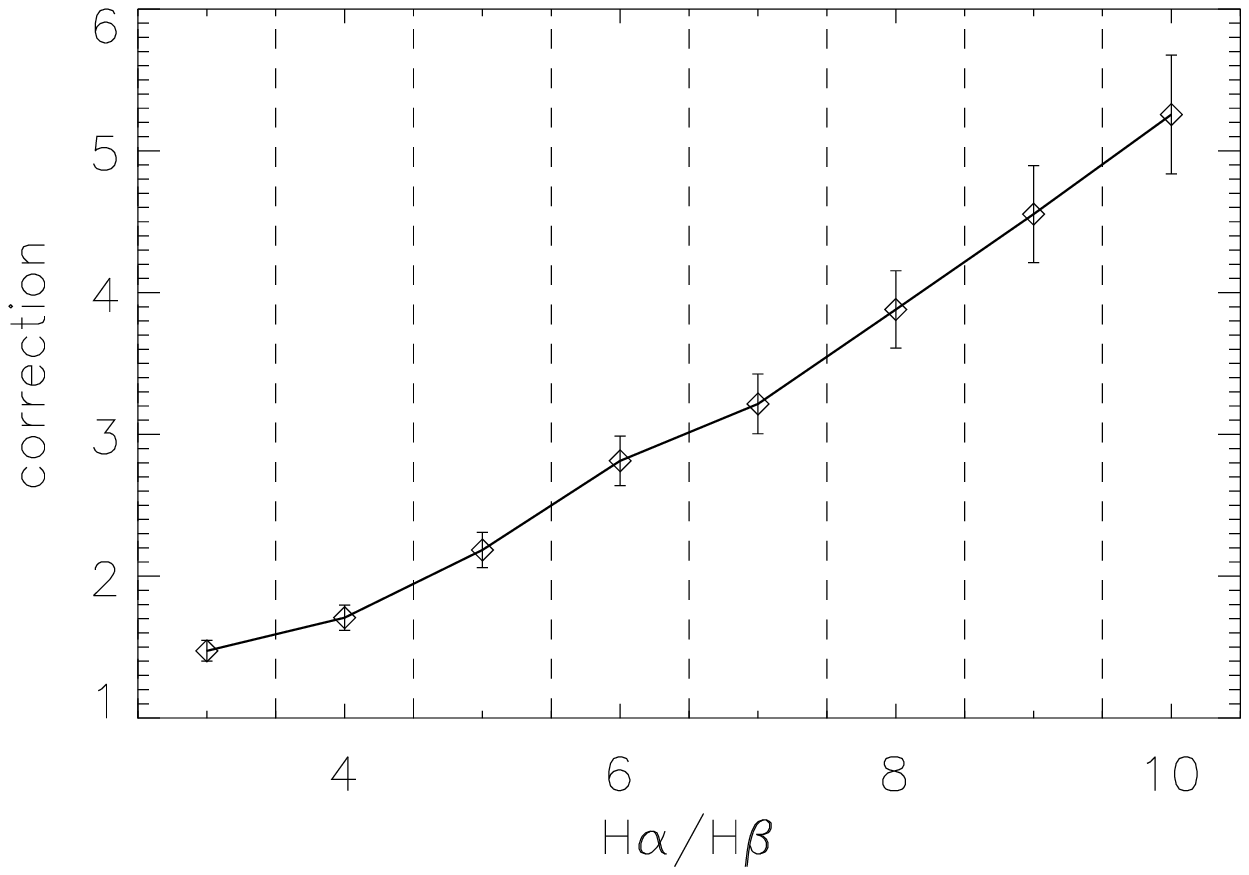}}
\put(40,-2){\includegraphics{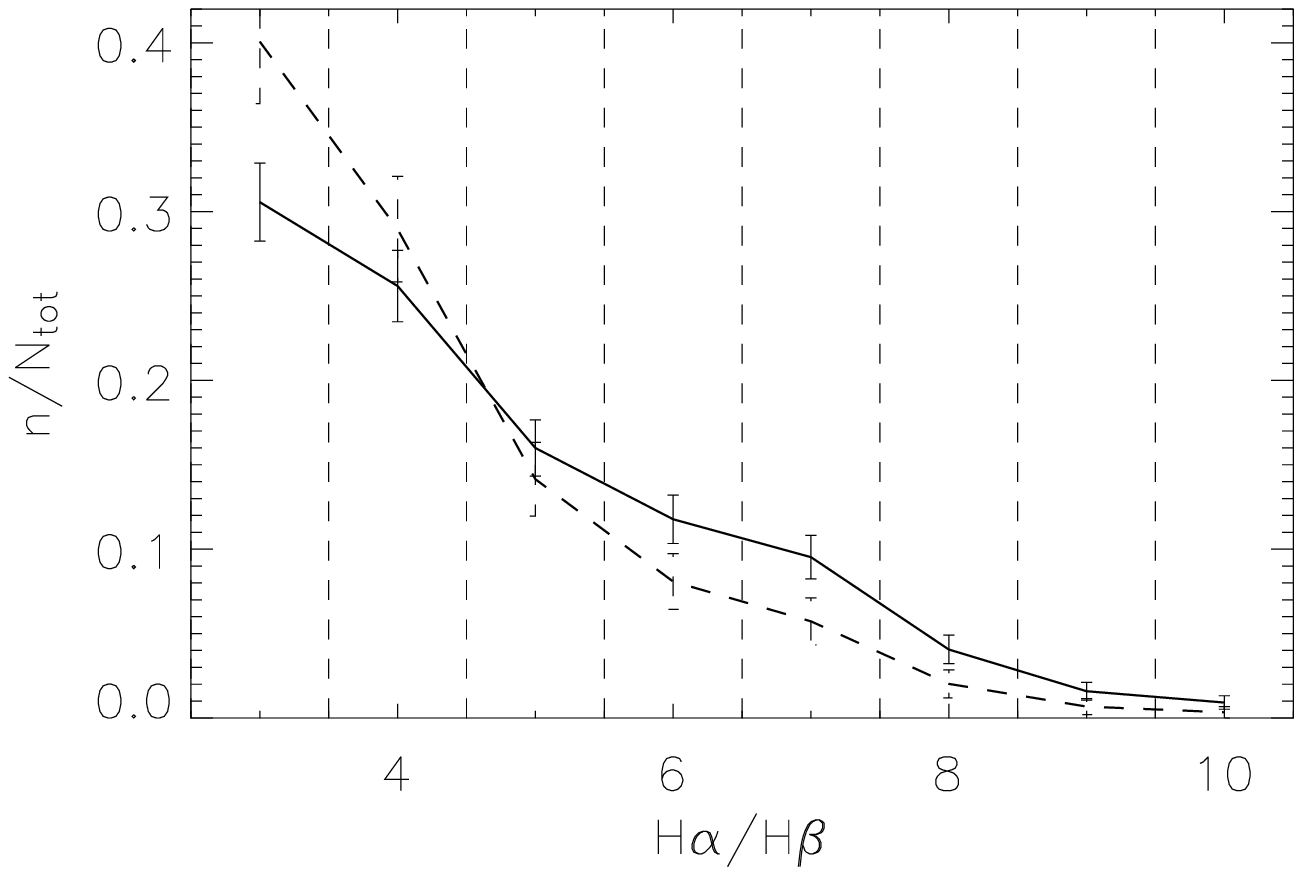}}
\end{picture}
\end{center} \caption{\label{fig:Ebv_correction}
Left panel: average correction factor
$N^{id}(E_{B-V})/N^{sim}(E_{B-V})$ along Balmer decrement
(H$\alpha$/H$\beta)_{BL}$. Right panel: the observed
 (dashed line) and corrected (solid line) Balmer decrement distribution.}
\end{figure}

We measure the emission line parameters of these simulated spectra
and determine whether they are broad line AGN as for the real
spectra. In this way, for each object, we obtain a probability as a
function of the black hole mass, un-attenuated nuclear luminosity,
the amount of extinction, and apparent magnitude, i.e.,
$S(M_\bullet,L,E_{B-V}, L^s)$. With this selection function, we can
estimate approximately the distribution of $E_{B-V}$ from the
observed one by assuming that the real $E_{B-V}$ distribution does
not depend on the black hole mass and optical luminosity. Under this
assumption, we can correct the observed $E_{B-V}$ distribution using
the above simulated spectrum
\begin{equation}
p(E_{B-V})= p^{obs}(E_{B-V}) N^{id}(E_{B-V})/N^{sim}(E_{B-V}).
\end{equation}
where the $p^{obs}(E_{B-V})$ and $p(E_{B-V})$ are the observed  and
real distribution of $E_{B-V}$; $N^{id}(E_{B-V})$ and $N^{sim}(E_{B-
V})$ are the number of spectra identified as Seyfert 1 galaxies and
total number of the simulated spectra that meet the SDSS targeting
criteria. For this correction, we have assumed that $E_{B-V}$
distribution is independent of black hole mass and nuclear
luminosity. The assumption has not been fully tested, however, it
should not dramatically affect our result as far as the correction
remains small or the dependence is weak. The observed and corrected
H$\alpha$/H$\beta$ is displayed in Figure \ref{fig:Ebv_correction}.
In comparison with the observed one, which peaks at 3.1, and is
identical to the value reported by Dong et al. (2008), the corrected
H$\alpha$/H$\beta$ distribution shows much more objects at higher
extinctions. Integrating over the distribution of $E_{B-V}$, we
yield:
\begin{equation}
S(M_\bullet, L, L^s)=\int S(M_\bullet,L,E_{B-V}, L^s) p(E_{B-V}) dE_{(B-V)}
\end{equation}
Because the correction increases fast with the degree of extinction,
one must keep in mind that the true number may become less reliable
at higher extinctions. For this reason, we will limit our analysis
only to $E_{B-V}<1.3$ (i.e., H$\alpha$/H$\beta<10$), the same
criteria we adopted in the type 1 AGN selection.

For quasars, the conditional probability
$p(L^s|M_\bullet,L)=\delta(L^s-L)$ and $S(M_\bullet, L, L^s)$ is
independent of $L^s$. For the broad line objects selected from the
galaxy sample, there is a distribution of $L^s$ for a given nuclear
luminosity and black hole mass due to the significant contribution
from the host galaxy. Using the relation between the host galaxy and
the black hole for the type 2 Seyfert galaxies (see Figure
\ref{fig:MBH_Mr}) and the nuclear luminosity from
$L_{\rm{H\alpha,Br}}-L_{5100}$ relation (refer to Greene \& Ho
2007), we write

\begin{eqnarray}
\label{eqn:maxlike}
p(L^s|M_\bullet,L)=\frac{1}{2\pi\rm{\sigma}_1\rm{\sigma}_2}\int
dL^{host}
\exp\left(-\frac{(L^{host}-L^{host}(M_\bullet))^2}{2\rm{\sigma}_1^2}\right)\\
\nonumber
 \times\exp\left(-\frac{(L^s-L^{host}-L_{r}(L))^2}{2\rm{\sigma}_2^2}\right)
\end{eqnarray}

where $L^{host}(M_\bullet)$ is the average $r$-band host luminosity
in the AGN rest frame for a black hole mass $M_\bullet$, and
$\rm{\sigma}_1$ is the scatter of this relation; $L_r(L)$ is the
average $r$-band nuclear luminosity for a given $L$ which is
estimated via the broad H$\alpha$ luminosity. As we already seen,
$\rm{\sigma}_2$ is much smaller than $\rm{\sigma}_1$, so we can
approximate the second part of the integrand as a $\delta$ function.

We show contours of the average selection function on the black hole
mass versus nuclear luminosity plane in Figure
\ref{fig:selection_func}. The selection function strongly depends on
the BH mass and nuclear luminosity [\ion{O}{iii}]: the
incompleteness increases towards larger black hole mass and lower
luminosities. At [\ion{O}{iii}] luminosity below $10^{40.7}$ \ergs,
the selection function is below 0.2, even at black hole mass as low
as $10^8 M_{\sun}$. The dearth of type 1 AGN below this limit can be
well attributed to the selection effects.

\section{The Type 1 Fraction}

\begin{figure}
\begin{center}
\setlength{\unitlength}{1mm}
\begin{picture}(150,95)
\put(0,-6){\includegraphics{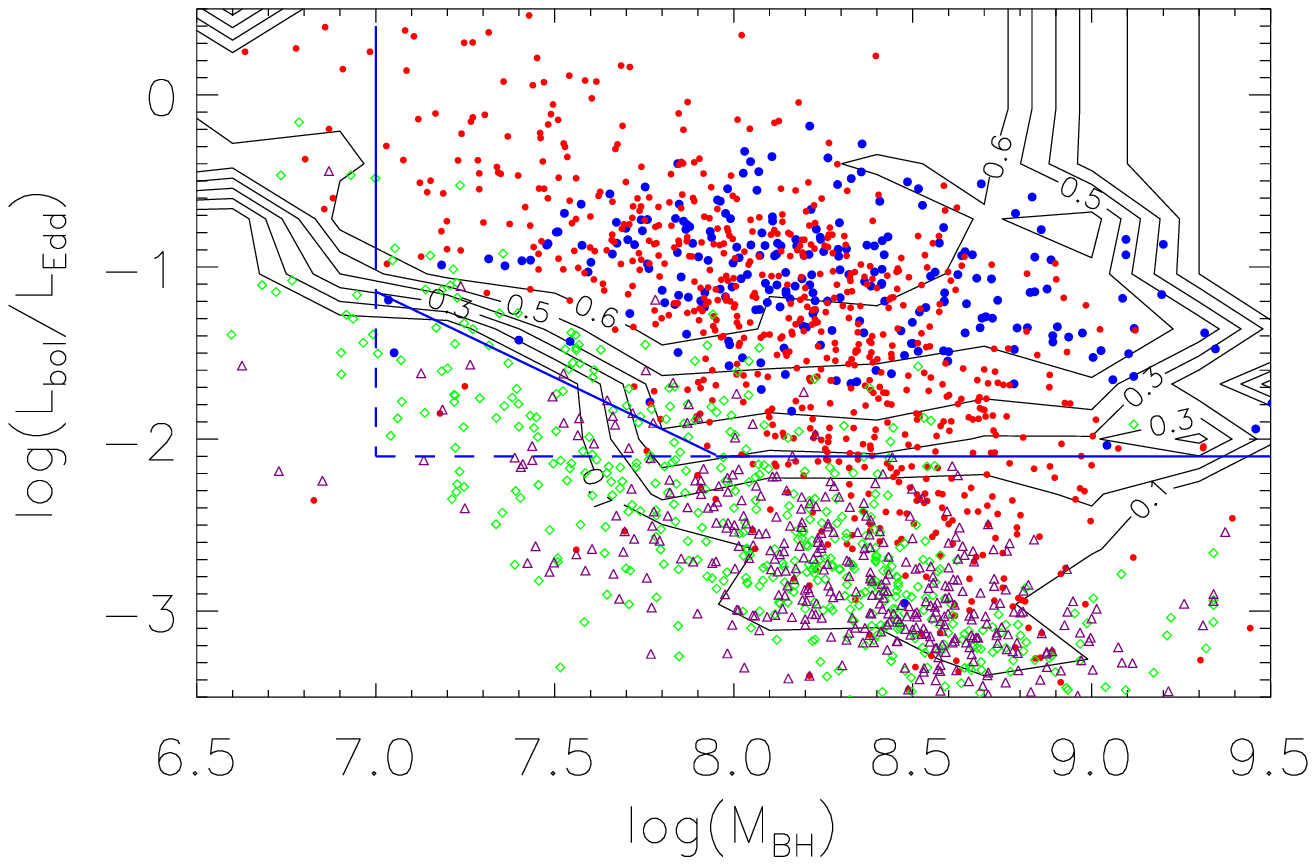}}
\put(0,45){\includegraphics{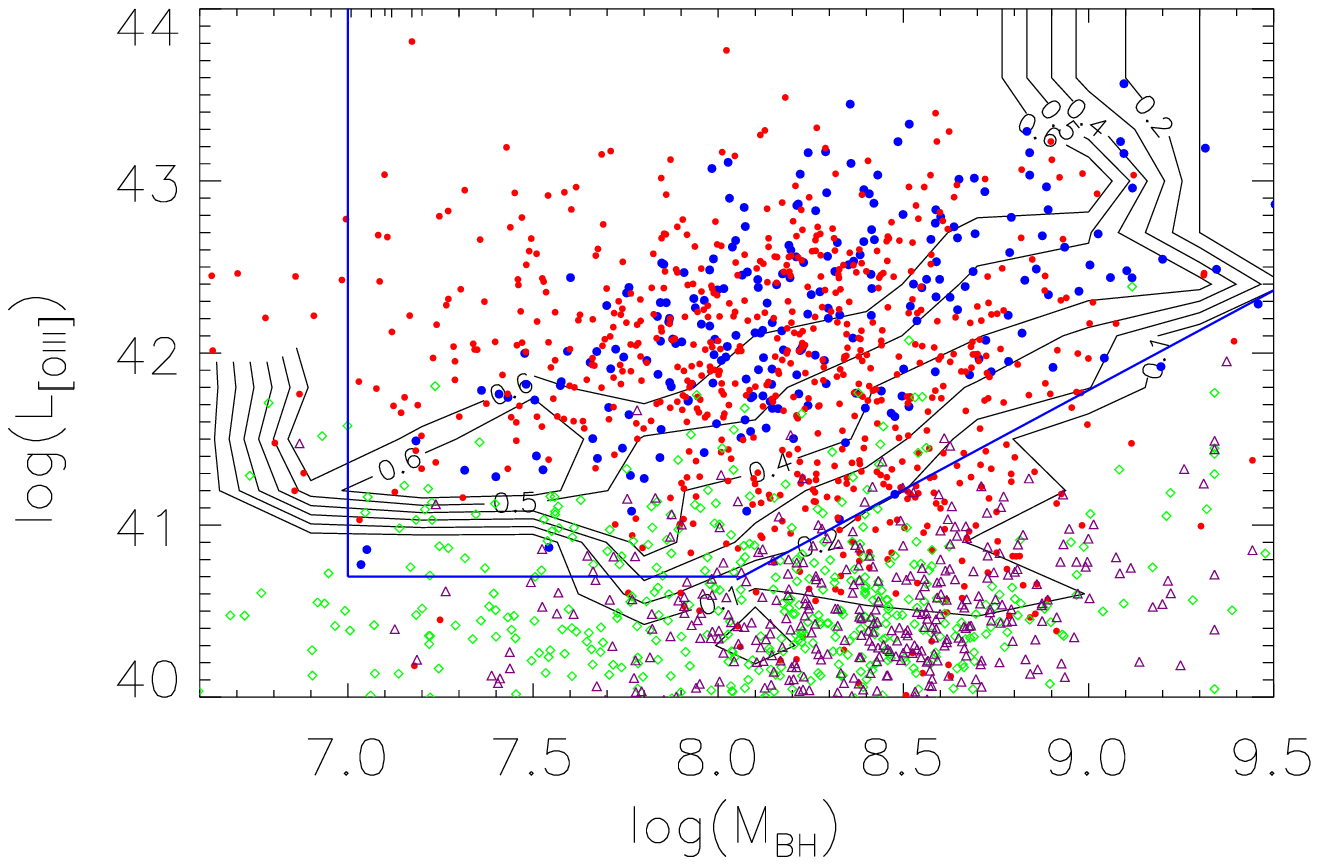}}
\end{picture}
\end{center} \caption{\label{fig:selection_func} The contour of selection function
 and the restricted regime of $\log(M_\bullet/M_{\sun})>6.6,\log(L_{\rm{[O III]}}\rm{/erg\ s^{-1}})>40.7$
 and $\log(L_{\rm{bol}}/L_{\rm{Edd}})>-2.1$ (blue lines). The red circles represent for type 2s,
 blue circles for type 1s, purple triangles for LINERs, and green diamonds for composite galaxies.
 The contour levels were set from 0.1 to 0.6, with interval of 0.1.}
\end{figure}

With all the above corrections, we will analyze the fraction of the
type 1 AGN as a function of the basic parameters, such as black hole
mass, accretion rate, nuclear luminosity and radio power. As we can
see in the Figure \ref{fig:selection_func}, due to selection effect,
the type 1 AGN can be detected to a reasonable fraction only in a
very limited parameter space on either $M_\bullet$ versus
$\log(L/L_{\rm{Edd}})$ plane or $L_{\rm{[O III]}}$ versus
$M_\bullet$ plane. In order to avoid potentially large uncertainties
introduced with our correction of selection effects, we will limit
our analysis to the parameter regimes of
$\log(M_\bullet/M_{\sun})>6.6$, $\log(L_{\rm{[OIII]}}\rm{/erg\
s^{-1}})>40.7$
 and $\log(L_{\rm{bol}}/L_{\rm{Edd}})>-2.1$, that a substantial fraction of Type 1 AGN are detected.
Using the selection function $S(M_\bullet, L, L^s)$, $V_{\rm{max}}$
and $p(L^s|M_\bullet, L)$ in the last section, we can calculate
$\phi(M_\bullet, L)$ for broad and narrow line AGN.

\subsection{The Dependence of Type 1 Fraction on $L_{\rm{[O III]}}$}

\begin{figure}
\begin{center}
\setlength{\unitlength}{1mm}
\begin{picture}(150,32)
\put(15,-6){\includegraphics{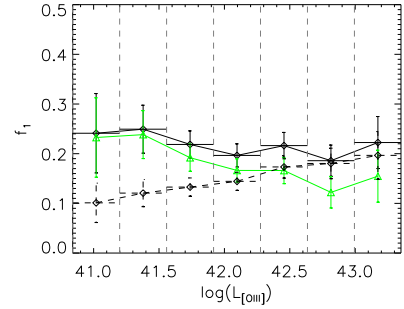}}
\end{picture}
\end{center}
{\caption[junk]{\label{fig:f1_LOIII} {The dependence of type 1
fraction on $L_{\rm{[OIII]}}$ for observed density (dashed line and
diamonds), bias corrected density(solid line and diamonds), and
maximum redshift evolution corrected density (green line and triangles). The
evolution correction is based on extrapolation of quasar luminosity
function of Richards et al. (2005). }}}
\end{figure}

The regime of $M_\bullet$ and $L_{\rm{[OIII]}}$ used in this
analysis is illustrated in the upper panel of Figure
\ref{fig:selection_func} with blue lines. In this regime,
$\log(L_{\rm{[OIII]}})$ distribution of type 1s and type 2s are very
similar with means $\log(L_{\rm{[OIII]}}\rm{/erg\ s^{-1}})$ of 42.47
and 42.40 for type 1 and type 2 AGN, respectively (see inserted
small diagram in Figure 10). We integrate $\phi(M_\bullet,
L_{\rm{[OIII]}})$ over either $M_\bullet$ in the corresponding range
to get $\phi(L_{\rm{[OIII]}})$ for both type 1 and type 2 AGN. It is
straight forward to calculate the type 1 fraction by dividing the
bias-corrected type 1 objects density by the total AGN density in
each bin. The results are show in Figure \ref{fig:f1_LOIII}.

The observed type 1 fraction increases with $L_{\rm{[OIII]}}$, which
is similar to, but with a flatter slope than those given by Simpson
(2005, hereafter S05) and Hao et al. (2005, hereafter H05) for radio
quiet AGN. After correcting for selection effects, $f_1$ keeps at
nearly a constant value of 20\% over the [\ion{O}{iii}] luminosity
range $40.7<\log(L_{\rm{[OIII]}}\rm{/erg\ s^{-1}})<43.5$. The least
$\chi^2$ fit for a constant $f_1$ yields $f_1\sim20.1\%$ which
is acceptable at a probability of 0.92.
This result is different from S05 and H05, who found that the fraction
of Type 1 AGN increases significantly with the nuclear luminosity for
mainly radio quiet AGN. The difference can not be considered (solely)
as a difference between radio-loud and radio quiet AGN, but be
attributed to several different treatments. First, S05 and H05 does not
corrected [\ion{O}{iii}]
luminosity for the internal extinction, while such a correction is
included in this work. It is known that type 2 Seyfert galaxies show
systematically larger Balmer decrements of narrow lines than Type 1
Seyfert galaxies (Cohen 1983; Gaskell 1984; Rhee \& Larkin 2005).
This extinction would reduce the [\ion{O}{iii}] luminosities of
Seyfert 2 galaxies with respect to the Seyfert 1 galaxies
systematically. In fact, before extinction correction, the type 1
fraction in our sample would also increase dramatically from $\sim
10\%$ at $\log(L_{\rm{[OIII]}}\rm{/erg\ s^{-1}})\sim 41$ to $\sim
90\%$ at $\log(L_{\rm{[OIII]}}\rm{/erg\ s^{-1}})\sim 43$. Second,
Type 2 AGN in the S05 includes also LINERs and composite type AGN.
While it is still controversy whether composite type 2 AGN are of
similar nature as Seyfert galaxies, those spectroscopic LINERs are
certainly different from Seyfert galaxies (Kauffmann et al. 2003;
Kewley et al. 2006; Heckmann et al. 2004). Also we note that most
composite galaxies are located in relative lower black hole mass and
LINERS on lower accretion rate regime, while broad lined AGN seldom
fall in these regions (refer to Figure \ref{fig:selection_func}).
Third, as seen in Figure \ref{fig:selection_func}, the difference
can be attributed at least partly to that we focus our analysis to
the limited parameter regime. Outside the regime, type 2 AGN are
detected in large amount but almost no type 1 object has been
detected on the regime with large black hole mass and small
[\ion{O}{iii}] luminosity. Finally, S05 did not consider selection
effect at all while H05 used a different selection function.
However, as seen in the Figure \ref{fig:selection_func}, this
changes the [\ion{O}{iii}] luminosity dependence only moderate. This
is understandable because we limit our analysis to the parameter
regime, where the correction is only modest.

Hitherto we ignored the redshift evolution, but $V/V_{max}$ test
does show mild positive evolution with $z$. We estimate the maximal
impact of the evolution on our result by using the double power-law
form quasar luminosity function $\rm{\Phi}(L,z)$ for $z<2.1$ derived
from 2dF-SDSS LRG and quasar survey (Richards et al. 2005). Assuming
that type 1s and type 2s evolved in the same form, in every
$\{L,z\}$ bin, we can apply correction of
$\rm{\Phi}(L,z=0)/\rm{\Phi}(L,z)$ to both type 1s and type 2s. The
result is shown in figure \ref{fig:f1_LOIII} (green line and
triangles). The redshift evolution pulls down the type 1 fraction by
$\sim6$ percent at high luminosity end, but change little at low
luminosity end. As expected, this effect does not significantly
alter our result.

\subsection{The Dependence of Type 1 Fraction on $M_\bullet$ and Eddington ratio}
\begin{figure}
\begin{center}
\setlength{\unitlength}{1mm}
\begin{picture}(150,32)
\put(15,-7){\includegraphics{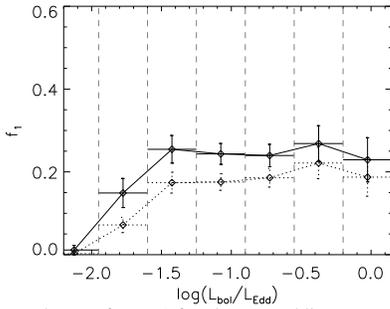}}
\end{picture}
\end{center}
{\caption[junk]{\label{fig:f1_Eddington_ratio} {The dependence of
type 1 fraction on Eddington ratio $\log(\ell)=\log(L/L_{\rm{Edd}})$
for observed density (dashed line) and bias corrected density (solid
line).} }}
\end{figure}

\begin{figure}
\begin{center}
\setlength{\unitlength}{1mm}
\begin{picture}(150,30)
\put(15,-6){\includegraphics{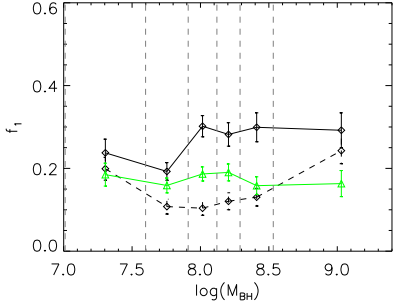}}
\end{picture}
\end{center}
{\caption[junk]{\label{fig:f1_MBH} The dependence of type 1 fraction
on $M_{\bullet}$ for the bias corrected density ($f_1^{\rm{cor}}$,
solid black line and diamonds), the observed density
($f_1^{\rm{obs}}$, dashed black line and diamonds), and simulated
type 1 density ($f_1^{\rm{sim}}$, green line and triangles, based on
the convolved type 2 density). Each bin contains the same number of
sources. }}
\end{figure}

The regimes of $M_\bullet$ and $\ell=L_{\rm{bol}}/L_{\rm{Edd}}$ used
in this analysis are shown in the bottom panel of Figure
\ref{fig:selection_func}. In this regime, the average $M_\bullet$
for type 1s ($\sim 10^{8.04}M_{\sun}$) and type 2s ($\sim
10^{8.11}M_{\sun}$) are indistinguishable (see small diagram in
Figure \ref{fig:MBH}). Within the regime, we integrate
$\phi(L,M_\bullet)$ over either $L$ in the corresponding range to
get $\phi(M_\bullet)$. Similarly, by integrating of
$\varphi(M_\bullet,\ell)$ (see Eq.\ref{eq:phi_M_lambda}) over
$M_\bullet$, we can obtain $\varphi(\ell)$ for both type 1 and type
2 AGN. Note that our analysis is restricted to the parameter range
illustrated in Figure \ref{fig:selection_func}(blue lines), where a
substantial fraction of both type 1 and type 2 AGN are detected.
Thus we can calculate the type 1 fraction straightly by dividing the
bias-corrected type 1s density by the total AGN density in each
$\log (\ell)$ bin and $\log(M_\bullet)$ bin. The results are show in
Figure \ref{fig:f1_Eddington_ratio} and Figure \ref{fig:f1_MBH}.

From Figure \ref{fig:f1_Eddington_ratio}, we find that the type 1
fraction keeps nearly a constant value of 25\% when Eddington ratio
changes from $\log(\ell)\simeq -1.5$ to $\log(\ell) \simeq 0.0$ for
the black hole mass in the range of
$6.6<\log(M_\bullet/M_{\sun})<9.5$. We also note that the correction
to selection effects is only modest in whole range of Eddington
ratio, and does not significantly affect our conclusion as shown in
the Figure \ref{fig:selection_func}. In fact, before the correction
for selection effects, type 1 fraction is more consistent with a
constant value over $-1.5<\log(\ell)<0.0$. At lower Eddington
ratios, type 1 fraction becomes very low, and the transition occurs
at $\log(\ell)\sim -2$. The transition is likely real, rather than
caused by under-estimate of the selection function. As seen in
Figure \ref{fig:selection_func}, most detected objects in this
region are massive objects, and the value of the selection function
is between 0.3 and 0.5. Lack broad emission line of those objects
maybe represent a true difference in the BLR properties or
over-luminous of [\ion{O}{iii}] in comparison with its true nuclear
luminosity in this parameter region.

As mentioned in \S 2.4, the uncertainty in the black hole estimate
will broaden the distribution, thus affect our results about the
dependence of type 1 fraction on black hole mass. Black hole masses
for type 2 are estimated using the $M_\bullet-\sigma_*$ relation.
This relation has an intrinsic scatter of only 0.3 dex, while the
Type 1 BH masses are estimated using the Virial estimator, which may
affected by the inclination of the AGN and accretion rate, and has a
typical uncertainty of a factor of 0.5 dex. Because this larger
scatter, a tail to high masses and low mass would be greater for the
Type 1s than the Type 2s when cut-off (such as $L_{\rm{[OIII]}}$) is
applied. When a ratio of Type 1 over Type 2 is taken, this would
then give an increasing Type-1 fraction at high and low black hole
masses as observed.

In order to correct this bias, we convolve the type 2
$\log(M_\bullet)$ distribution with a Gaussian of $\sigma=0.4$, so
that type 1s and type 2s have the same scatter. We restrict to
objects in \{$L_{\rm{[OIII]}}, M_\bullet$\} region, as shown in
upper panel of Figure \ref{fig:selection_func}. Based on convolved
type 2 density $\rho^{\rm{conv}}_{\rm{type2}}$, the simulated type 1
density fraction $f_1^{\rm{sim}}=
 \rho^{\rm{obs}}_{\rm{type}}/(\rho^{\rm{obs}}_{\rm{type1}}+\rho^{\rm{conv}}_{\rm{type2}})$
 was displayed in Figure \ref{fig:f1_MBH}, represented with green line and triangles.
  While the observed type 1 density fraction
$f_1^{\rm{obs}}=
\rho^{\rm{obs}}_{\rm{type1}}/(\rho^{\rm{obs}}_{\rm{type1}}+\rho^{\rm{obs}}_{\rm{type2}})$
 was represented with dashed line and diamonds. Therefore, in every $\log(M_\bullet)$ bin, we could
correct the selection effect from different $M_\bullet$ measurement
uncertainty for a factor of $f_1^{\rm{sim}}/f_1^{\rm{obs}}$.
Combining it with selection function and detection probability, we
plot the type 1 fraction $f_1^{\rm{cor}}=\phi_1/(\phi_1+\phi_2)$
 dependence on $\log(M_\bullet)$ in figure \ref{fig:f1_MBH} with solid line.
The least $\chi^2$ fit for a constant $f_1$ yields
$\chi^2=15.54$ for 5 degrees of freedom, which rules out a constant
$f_1$ at a confidence level of $\sim 99\%$.
 From the figure,
 we can see that the type 1 fraction is $\sim30\%$ in higher BH mass
 region ($\log(M_\bullet/M_{\sun})>8$),
  about 10 percent higher
 than that in lower $M_\bullet$ region.
$f_1$ also rises in the BH mass bin below
$\log(M_\bullet/M_{\sun})<7.6$. This may be due to the loss of type
2 AGN in this sample. Low mass black holes are usually hosted in the
disk galaxies with relative small bugle, such galaxies are more
likely to have a star forming disk. Due to relative large aperture
of SDSS fibre, SDSS spectrum will encompass also the star forming
disk. This will shift some type 2 Seyfert galaxies into the
composite-type. Because we consider only Seyfert 2 type spectra,
those type 2 AGN will be missed in our Seyfert 2 sample (refer to
the ``composite galaxies'' in Figure \ref{fig:selection_func},
represented with green points). In order to check the fraction of
such objects, we examine the distribution on the BPT diagram of type
1 AGN, which are selected based solely on the presence of broad
lines. We find that only 4.05\%(9) type 1s with
$\log(M_\bullet/M_{\sun})>7.6$ show composite spectra on BPT
diagram, while 21.3\%(16) of broad line AGN with
$\log(M_\bullet/M_{\sun})<7.6$ do. This confirms our suspicion that
the contamination of star-formation region becomes more important at
low black hole masses.

\subsection{The Dependence of Type 1 Fraction on Radio Properties}

\begin{figure}
\begin{center}
\setlength{\unitlength}{1mm}
\begin{picture}(150,35)
\put(-6,-2){\includegraphics{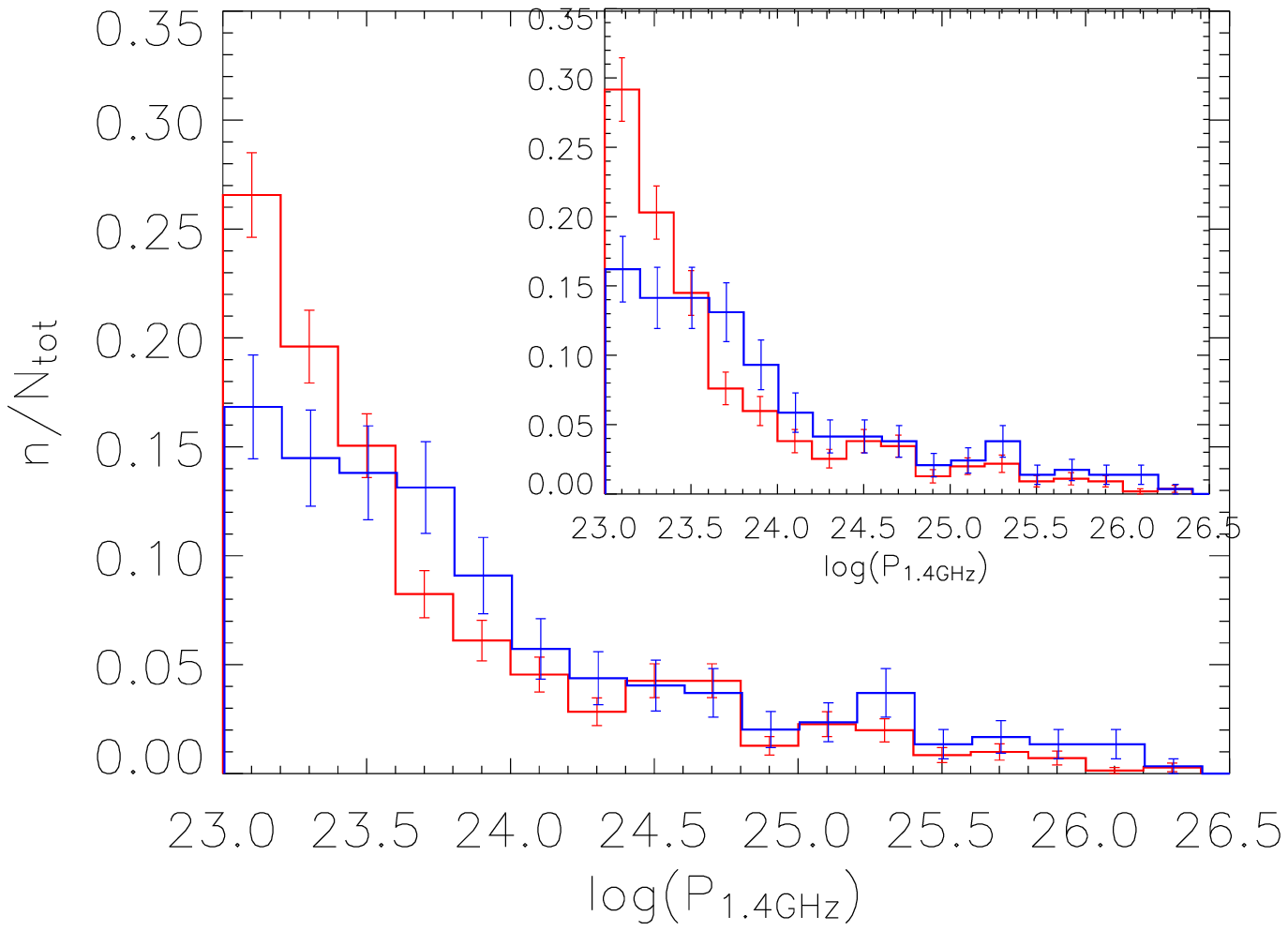}} \put(40,-2){\includegraphics{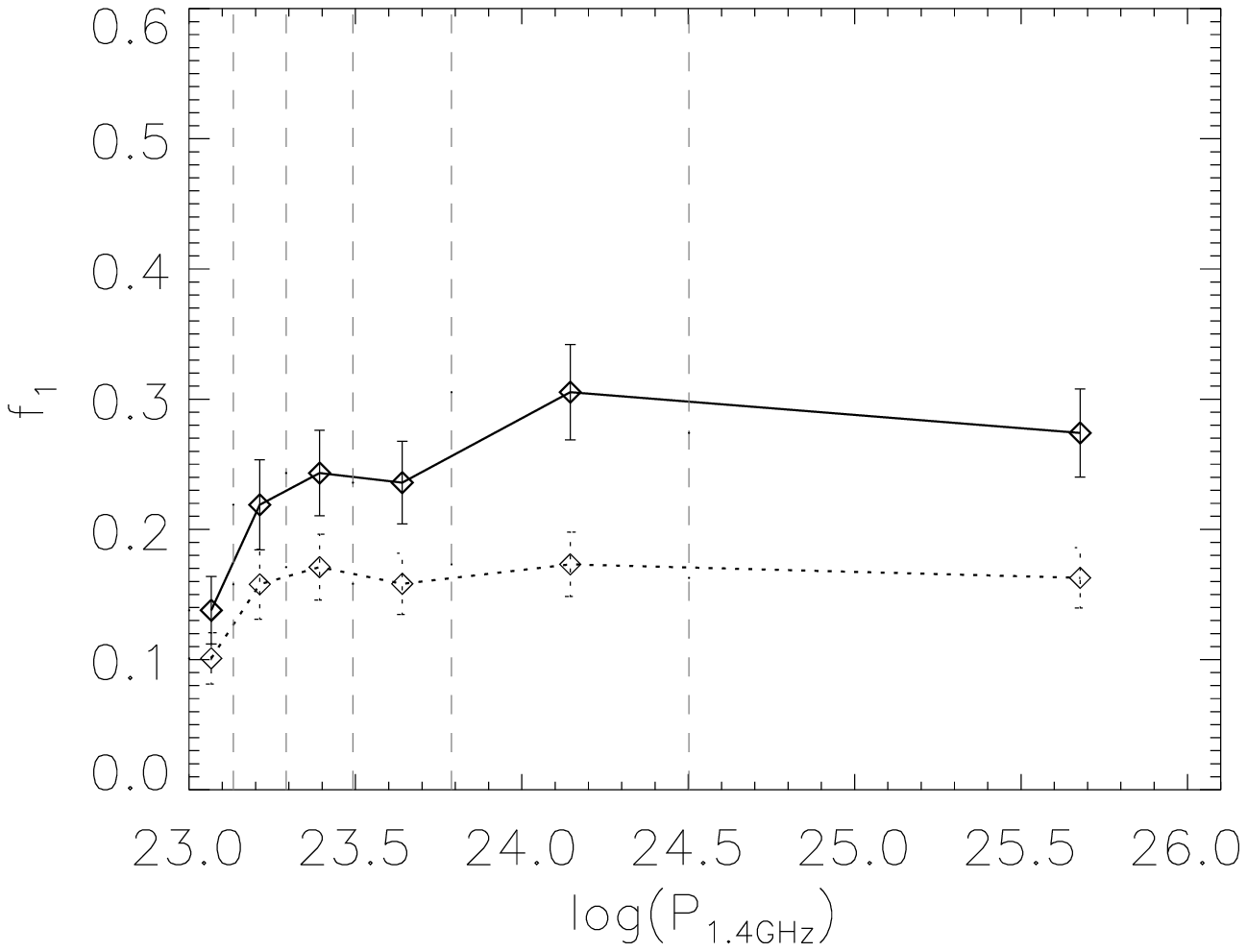}}
\end{picture}
\end{center}
{\caption[junk]{\label{fig:f1_L1.4GHz} {Left panel: the radio
power $P_{\rm{1.4GHz}}$ distribution for type 1s (blue line)
and type 2s (red line). The small diagram shows $P_{\rm{1.4GHz}}$
distribution within restricted $\log(L/L_{\rm{Edd}})$ and
$\log(M_\bullet)$ regime, refer to Figure \ref{fig:selection_func}.
Right panel: the dependence of type 1 fraction on $P_{\rm{1.4GHz}}$
for observed density (dashed line) and bias corrected density (solid
line). Each bin in right panel contains the same number of sources.}
}}
\end{figure}

We plotted the distribution of radio luminosity $P_{\rm{1.4GHz}}$ in
the left panel of Figure \ref{fig:f1_L1.4GHz}. It is clear that more
type 2 AGN distribute near the lower $P_{\rm{1.4GHz}}$ limit
$10^{23}$ W~Hz$^{-1}$ than type 1 objects, and fewer type 2 objects
in the high radio luminosity region $P_{\rm{1.4GHz}}>10^{25}$
W~Hz$^{-1}$. After limiting to radio AGN in the parameter regime
defined in Figure \ref{fig:selection_func}, the overall distribution
is also quite similar. The type 1 fraction increases from 15\% at
$\log (P_{\rm{1.4GHz}}\rm{/W\ Hz^{-1}})=23$ to nearly 30\% at $\log
(P_{\rm{1.4GHz}}\rm{/W\ Hz^{-1}})=24$ and then flattened to $\log
(P_{\rm{1.4GHz}}\rm{/W\ Hz^{-1}})=26$ (Figure \ref{fig:f1_L1.4GHz}).
Note that the trend is very similar before and after correction for
selection effects. $\chi^2$-test rules out the possibility of a
constant $f_1$ at a confidence level of 99.5\%. However, when the
lowest radio power bin is removed, the trend in the rest 5 bins is
consistent with a constant $f_1$ at a probability of 50\%. Lawrence
et al. (1991) found that type 2 fraction decreases with radio power
for a lower frequency selected 3CR radio sample, which consists of
mainly powerful radio sources. Since our sample includes only a
smaller number of powerful ($P_{\rm{1.4GHz}}>10^{26}$ W Hz$^{-1}$)
radio sources, thus is not adequate to address whether $f_1$ rises
at high radio luminosity.

Comparing to lower frequency selected samples, our sample is more
likely affected by radio selection effects. It is general considered
that radio emission from an AGN consists of two components, an
isotropic extended lobe component and a beamed jet component (e.g.,
Urry \& Padovani 1995). When the AGN is viewed along the radio jet,
the jet component with a relative flat radio spectrum is boosted
relativistically, while the lobe emission with a steep spectrum
remains the same, i.e., high frequency radio flux is boosted along
the jet direction more than low frequency radio flux. As such, a
high frequency survey is more sensitive to the core-dominated or
compact, flat-spectrum sources than to the lobe-dominated or
diffused, steep-spectrum sources. Because it is general believed
that the jet emerges along the symmetric axis, the sample will be
biased to the face-on system, whose optical spectrum is type 1 in
unified scheme.

The beaming effect can be checked with radio morphology. A radio AGN
will appear as a double-lobe source when it is observed side-away,
and as core-jet or core dominated structure viewed along the jet
direction due to beaming effect of the relativistic jet. Therefore,
the core dominance parameter, defined as $C=P_{c}/P_{t}$
($P_{t}=P_{c}+P_{l}$), as a rough indicator of the system
inclination. If radio jet aligns with the symmetric axis of the
dusty torus, one would expect that type 1 fraction increases with
$C$. However, only about 12\% sources in our sample have resolved
radio structures. The number is still too small to allow to reach a
firm conclusion on this.

\section{SUMMARY AND DISCUSSION}
We have quantified several selection effects on the AGN
classification and spectral targeting for different combinations of
host galaxy properties, nuclear properties and dust extinction from
spectroscopic samples of radio loud SDSS galaxies and quasars using
Monte-Carlo simulations. Type 1 AGN have much strong selection
effect than type 2 AGN. Type 1 radio AGN can be studied only in a
very limited parameter space on the black hole mass versus Eddington
ratio or luminosity diagram. After correction for these selection
effects, we find that in the limited parameter region: (1) type 1
fraction is nearly independent of extinction-corrected
[\ion{O}{iii}] luminosity and Eddington ratio at $\log
(L_{\rm{[OIII]}}\rm{/erg\ s^{-1}})>40.7$ and $\log(\ell)>-1.5$; at
very low Eddington ratio, there are almost no type 1 AGN, which can
not be accounted for solely by selection effects, indicating a
transition in the broad line strength around $\log(\ell)\sim-2$ (see
also Hopkins et al. 2009). (2) The type 1 fraction increases
with the radio power from $10^{23}$ to $10^{24}$ W Hz$^{-1}$, then
keep flat until $10^{26}$ W Hz$^{-1}$. (3) The type 1 fraction is
$\sim 30\%$ in $\rm{M}_\bullet>10^{8}M_{\sun}$ region, $\sim 10\%$
higher than that in lower $\rm{M}_\bullet$ region.

Our result on the luminosity dependence is quite different from
previous studies for radio loud quasar and galaxies (Lawrence 1991;
Hill et al. 1996; Reyes et al. 2008), or mainly radio quiet AGN from
SDSS (Simpson 2005, Hao et al. 2005). The difference can be mainly
attributed to three factors. First, we use an extinction corrected
[\ion{O}{iii}] luminosity while previous authors did NOT. The
average Balmer decrements of narrow components for Seyfert 2
galaxies ($\sim 6.08$) is significantly higher than for Seyfert 1
galaxies ($\sim 3.59$), which will give arise to a factor of 5
difference in [\ion{O}{iii}] luminosity for two type of Seyfert
galaxies. Second, we restricted our analysis to the parameter
regimes, where a substantial fraction of objects can be detected
according to selection function, this reduces the uncertainty caused
by the correction of selection effects. The final parameter regime
used here is similar to the one that found by Hopkins et al. (2009)
based on a different approach. This cuts down the objects with very
low Eddington ratios and high black hole mass, thus raises the type
1 fraction in the low luminosity end. Finally, our AGN
classification is based on more recent work of Kewley et al. (2006),
which gives a better separation of LINERs and Seyfert galaxies. The
new criteria removes quite some more LINERs than the previous
criteria. Because most of these LINERs have lower [\ion{O}{iii}]
luminosity (see Figure \ref{fig:selection_func}), this raises
significantly the fraction of type 1 in the low luminosity.

Although the X-ray observed AGN sample do not suffer from above
selection effects, however, the fraction of type 1 object based on
the optical classification into broad and narrow lined objects
suffers from the aforementioned selection effects. Thus it is
probably the same reasons caused a luminosity-dependence of obscured
type-1 (Hasinger et al.2008; Gilli et al. 2007; Fiore et al. 2008).
Rowan-Robinson et al. (2009) using infrared to X-ray spectral energy
distribution as a classification for obscured and un-obscured AGN,
and they reached a similar conclusion as this paper.

Two very different phonometrical models have been proposed for the
unification of broad and narrow line radio AGN. The receding torus
model was first proposed by Lawrence (1991) to explain the decrease
of the fraction of narrow line objects with the optical luminosity
in radio loud AGN, and was extended to radio quiet objects later
(Hill et al. 1996; Simpson 2005; Hao et al. 2005; Suganuma et al.
2006). The basic idea is that the opening angle of the dust torus is
larger for more luminous AGN because dust sublimation radius
increases with nuclear luminosity while the height of the torus is
assumed to be independent of the bolometric luminosity. Thus the
broad-line region can be seen over a larger opening angle in more
luminous objects. On the other hand, Grimes et al. (2004) showed
that observations are consistent with dual population radio sources,
'starved' low luminosity AGN, which do not have a broad emission
line region, and 'Eddington-tuned' high luminosity AGN, without
invoking a receding model. In our analysis, we excluded these very
low Eddington ratio objects \footnote{Most of low Eddington ratio
objects (with $L/L_{\rm{Edd}}<0.01$) are LINERs (see Figure
 \ref{fig:selection_func}).}. Our results show that the torus opening
angle does not change with the nuclear luminosity, thus do not
support 'receding torus' model, but are consistent with the two
population models.

Very little has been known about the dependence of the type 1
fraction on black hole mass or accretion rate. Using the ratio of
infrared to bolometric luminosity as an indicator of subtending
angle of torus, Cao (2005) found that the opening angle of torus
increases with the central black hole for a sample of Palomer-Green
quasars, but does not correlate with its Eddington ratio. On the
other hand, Zhou \& Wang (2005) argued that the subtending angle of
torus decreases with increasing accretion rate based on the
equivalent width of narrow FeK$\alpha$ line. It should be noted that
the origin of the narrow FeK$\alpha$ is not clear and the equivalent
width is also sensitive to the column density of the absorbing
material. Our results agree with Cao et al. (2005) in the region of
$\log(M_\bullet/M_{\sun})>7.6$.

It is difficult to compare our results with theoretical models of
dusty torus. Most dynamic-based dusty torus models are not able to
predict quantitatively how the opening angle of torus changes with
the black hole mass or Eddington ratio, although some relation may
be expected. In the dusty cloud model, the physical process that
maintains a thick torus is still not clear (Krolik \& Begelman 1988;
Zier \& Biermann 2002). Beckert \& Duschl (2004) proposed that a
geometrically thick torus can be sustained at a high accretion rate
by the balance of energy dissipation due to cloud-cloud collision
and heating due to accretion. They found a torus height to radius
ratio $H/R\sim\sqrt{2G\dot{M}/(c_sM(R))}$, where $\dot{M}$ is the
accretion rate, $M(R)$ the mass within radius $R$, and $c_s$ sound
speed. Their model has a clear prediction that subtending angle of
the torus increases with the accretion rate. Our results do not
support this. Radiation pressure support is suggested by Pier \&
Krolik (1992), and the infrared radiation and local heating pressure
can be very effective in supporting a smooth distributed torus.
However, an equilibrium solution can be found only in a relative
narrow Eddington ratio range (Krolik 2007; Shi \& Krolik 2008). It
has still to be demonstrated that models taken into consideration of
more physical processes, such as instability, would finally lead to
a prediction of a constant torus opening angle over a fairly broad
Eddington ratios.

In an alternative model of disk outflow, the entrained dusty cold
clouds are responsible for the obscuration (e.g., Ko¡§nigl \& Kartje
1994). Dopita et al. (1998) suggested that the torus and the
accretion disk may interacte by accretion-outflow feedback process.
Elitzur \& Shlosman (2006) argued that a torus is present only at
nuclear luminosity greater than 10$^{42}$ \ergs, and its covering
factor decreases with increase luminosity. However, its argument is
based on a strong assumption that the size of dusty clouds are
scaled only with their launching radius. On the other hand,
numerical simulations showed that luminous and high accretion rate
($L_{\rm{bol}}/L_{\rm{Edd}})$ AGN are likely strongly affected by
obscuration in the disk wind models (Schurch, Done \& Proga 2009).

\begin{figure}
\begin{center}
\setlength{\unitlength}{1mm}
\begin{picture}(150,35)
\put(-6,-2){\includegraphics{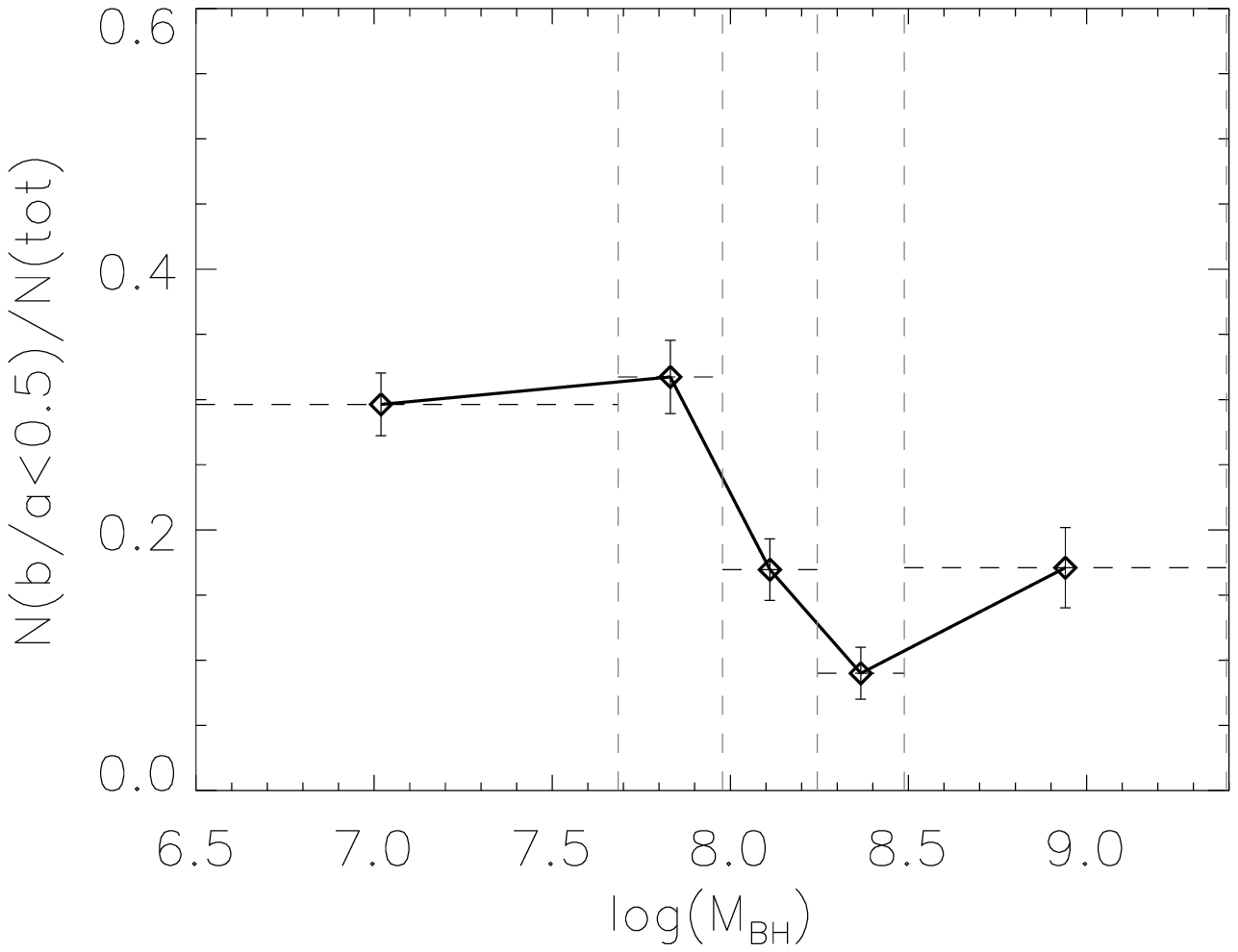}} \put(40,-2){\includegraphics{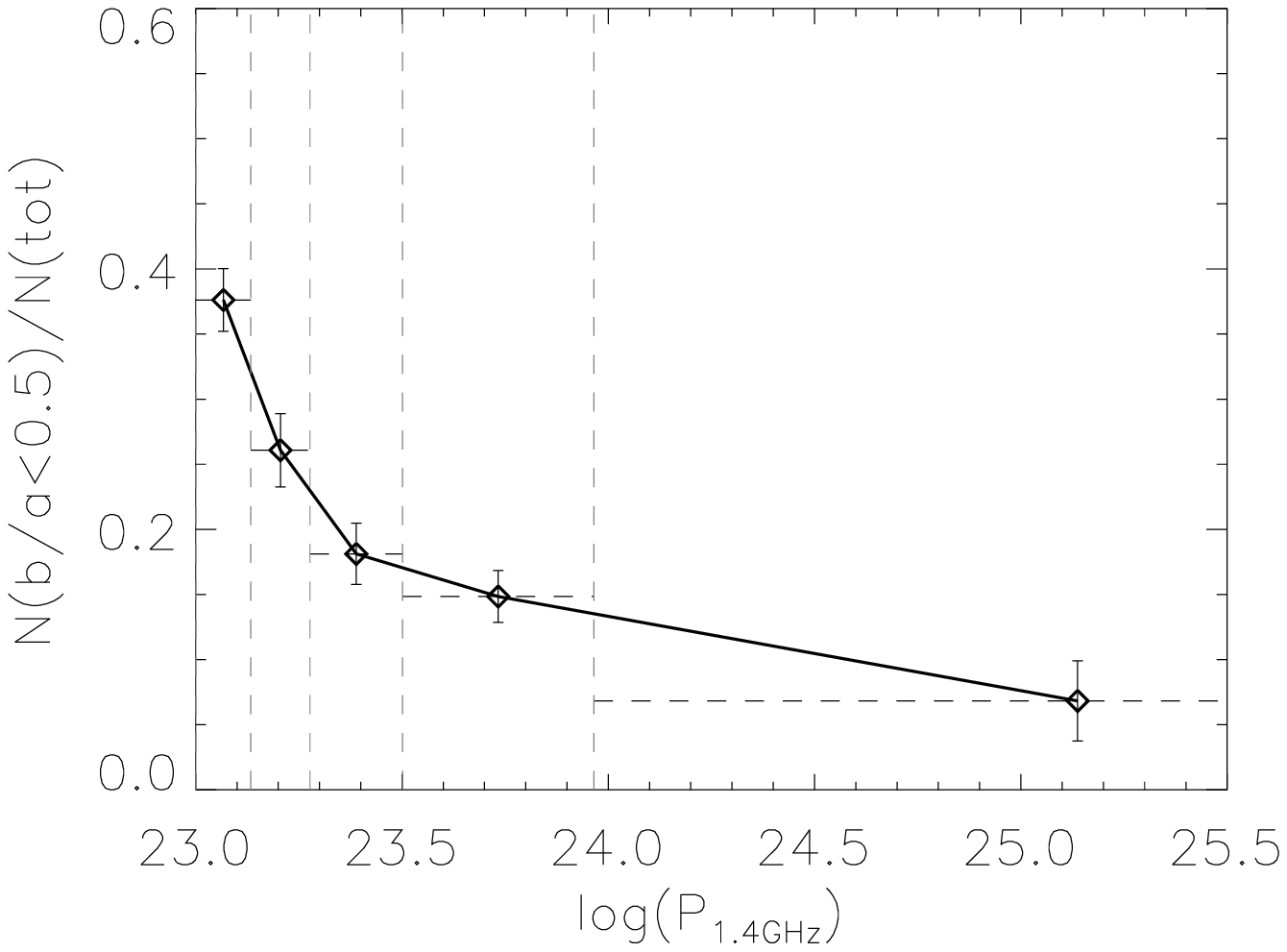}}
\end{picture}
\end{center}
{\caption[junk]{\label{fig:ab} { The fraction of high inclination
((b/a)$_{\rm{exp}}<0.5$) type 2s among disk host galaxies. Left
panel displayed its dependence on $\log(M_\bullet)$, and the right
panel displayed its dependence on $\log(P_{\rm{1.4GHz}})$. Each bin
contains the same number of sources.}}}
\end{figure}

It should be noted that there are obscuration sources other than
dusty torus. The large-scale galactic dust lane and the nuclear
star-burst region are certainly responsible for obscuration in some
of observed type-2 AGN. As we have noticed that a fraction of type-2
objects in the sample resides in relative edge-on disk galaxies. In
Fig \ref{fig:ab}, we displayed the fraction of high inclination (b/a
$<0.5$) type 2 galaxies among disk host galaxies. The fraction
 decreases quickly as black hole mass and radio luminosity increases.
This matches almost exactly the variations in $f_1$ with
$M_\bullet$ found in Fig \ref{fig:f1_MBH}: a $\sim10\%$ step
function in $M_\bullet$ at around $10^8 \rm{M_{\sun}}$. Also, the
decrease of the edge-on disks fraction in right panel of Fig
\ref{fig:ab} may be related to the large increase of $f_1$ in the
lowest radio power bin in Fig \ref{fig:f1_L1.4GHz}. These results
can be explained by the the presence of a population of Seyfert
galaxies, seen as Type-2 due to galaxy-scale obscuration. Therefore,
the increase of type 1 fraction with black hole mass and radio
luminosity may be understood as that host galaxies moves gradually
from disk to early-type galaxies, the extinction by the host galaxy
decreases. This result offer a very different interpretation from
the rest of the paper discusses.

\section{ACKNOWLEDGEMENT}
 We thank the anonymous referee for useful suggestions to
improve the paper. This work was supported by the Chinese NSF
through NSF-10973013 and 973 program 2007CB815403. This paper has
made use of the data from the SDSS and FIRST. Funding for the
creation and distribution of the SDSS Archive has been provided by
the Alfred P. Sloan Foundation, the Participating Institutions, the
National Aeronautics and Space Administration, the National Science
Foundation, the US Department of Energy, the Japanese Monbukagakusho
and the Max Planck Society. FIRST is funded by the National
Astronomy Observatory (NRAO), and is a research facility of the US
National Science foundation and uses the NRAO Very Large Array.

\label{lastpage}

\end{document}